\documentclass{optica-article}

\journal{opticajournal}

\articletype{Research Article}

\usepackage{lineno}
\usepackage[utf8]{inputenc}

\usepackage{makecell}

\begin{document}

\title{Superconducting single-photon detectors for integrated quantum photonics}

\author{Ilya A. Stepanov,\authormark{1} Oksana I. Shmonina,\authormark{1} Evgeniy V. Sergeev,\authormark{1,2} Aleksandr S. Baburin,\authormark{1,2}  Danila Yu. Ulyanov,\authormark{1} Kirill A. Buzaverov,\authormark{1,2} Sergey S. Avdeev,\authormark{1,2} Aleksey B. Kramarenko,\authormark{1} Yuri V. Panfilov,\authormark{1} and Ilya A. Rodionov\authormark{1,2,*}}

\address{\authormark{1}Shukhov Labs, Quantum Park, Bauman Moscow State Technical University, Moscow 105005, Russia\\
\authormark{2}Dukhov Automatics Research Institute (VNIIA), Moscow 127055, Russia\\}

\email{\authormark{*}irodionov@bmstu.ru}

\begin{abstract*} 
Single-photon detection possibility is a fundamental requirement for quantum technologies, including communication, computing and sensing. To achieve scalability and practical deployment, increasing attention is being directed toward integration of detectors with photonic integrated circuits, which offer compactness and compatibility with mass production. Superconducting nanowire single-photon detectors have emerged as the leading solution, combining near-unity efficiency, high temporal performance and the ability to be embedded across a wide range of photonic material platforms. In this review we trace the development of integrated superconducting nanowire single-photon detectors from early demonstrations to recent advances, outlining the progress in device architectures, material engineering and integration strategies. We also discuss performance benchmarks, emerging alternative designs, the future opportunities and challenges for this rapidly evolving field.

\end{abstract*}

\section{Introduction}
 Superconducting nanowire single-photon detectors (SNSPDs) have emerged over the past two decades as one of the most advanced technologies for single-photon detection owing to their unique combination of characteristics. Thanks to the ability to achieve system detection efficiency of up to 99\%\cite{chang2021detecting, reddy2019achieving, hu2020detecting}, count rate as high as 1 GHz\cite{zhang201916, resta2023gigahertz}, jitter less than 3 ps\cite{korzh2020demonstration} and dark count rate around $10^{-3}$ Hz\cite{shibata2015ultimate}, SNSPD finds applications in a variety of fields, providing breakthrough results in quantum computing\cite{deng2023gaussian, todaro2021state}, quantum communications\cite{liu2023experimental, liao2018satellite}, bioimaging\cite{tamimi2024deep, kim2025optical}, astronomy\cite{baudis2024first} and fundamental research\cite{liu2019experimental}.

However, as the field of scalable quantum technologies continues to mature, there is an increasing demand for integrating SNSPDs with other components on a single photonic chip, including low-loss waveguides\cite{buzaverov2023low}, single-photon sources\cite{bogdanov2018ultrabright} and electro-optical (EO) modulators\cite{lotkov2024integrated}. Such integration not only minimizes optical connection losses but also enhances system scalability, enabling multichannel detection in a single cryostat without compromising performance. Compared to other promising types of single-photon detectors, such as single-photon avalanche diodes (SPADs) and transition-edge sensors (TESs), SNSPDs not only exhibit comprehensively superior performance in standalone configurations but also offer the simplest technological path for waveguide integration\cite{wang2025advances}. In contrast, monolithic integration for other single-photon detector types faces significant technological challenges, particularly in achieving sensitivity at telecom wavelengths and maintaining operation with photonic integrated circuit elements at millikelvin temperatures, and has not yet been widely demonstrated to date\cite{calkins2013high, martinez2017single, zhu2017all, xiang2022high}.

Several recent review articles have covered the development of standalone superconducting single-photon detectors, discussing their operational principles and state-of-the-art metrics\cite{esmaeil2021superconducting}, superconducting materials used\cite{shibata2021review,tripathy2024comparative}, research trends\cite{venza2025research, lau2023superconducting, taylor2026mid} and scaling\cite{gao2025pixels}. In contrast, integrated SNSPDs (iSNSPDs) have received less attention despite the numerous unresolved challenges in this field. Although several review articles have addressed progress in iSNSPD development and their applications \cite{ferrari2018waveguide, kovalyuk2024waveguide, raj2025waveguide}, the field has evolved rapidly, with significant advances made in recent years. Consequently, several emerging architectures and fundamental aspects of device physics have not yet been thoroughly examined in a consolidated work. Thus, a comprehensive and up-to-date review is needed to summarize the current state of integrated SNSPDs, focusing on practical performance metrics, the challenges and opportunities in large-scale integration into high-performance systems for quantum information processing. 

In this review, we examine recent advances in integrated SNSPDs, detailing their performance metrics, architectures, measurement techniques, materials used and demonstrated characteristics. In Section~2, we describe the working principle of SNSPD and review the main theoretical models devoted to the description of the microscopic processes occurring during detection, highlighting the limitations of existing models and assessing their consistency with experimental observations. Section~3 examines both the performance metrics of iSNSPDs and system parameters that influence them, including material properties, detector geometry, operating conditions and light parameters, thereby outlining the parameter space available for detector performance optimization. Section~4 evaluates superconducting materials and photonic platforms used in iSNSPDs, along with the various approaches to their integration, and compares their advantages and limitations, while providing a comprehensive overview of the performance of all reported devices across different waveguide–superconductor combinations. Section~5 reviews current approaches for implementing photon-number resolution in integrated SNSPDs, examining their distinctive operating principles and inherent limitations. Section~6 provides a forward-looking perspective on outstanding challenges and emerging opportunities in the ongoing development of integrated superconducting single-photon detectors. Finally, Section~7 summarizes the main results of this review and outlines the current state of iSNSPDs development.

\section{Working principle}

Numerous experimental and theoretical studies are currently devoted to investigating SNSPD detection mechanisms and its response modeling\cite{haldar2024modeling, he2025unified, mcnaughton2023consequences, dane2022self, bauer2025type, simon2025ab}. It is worth noting that the theoretical models describing detection mechanisms remain universally applicable to both stand-alone and integrated SNSPDs. While a complete understanding of these processes is still lacking, comprehensive analysis of the phenomena that occur in the superconducting strip following photon absorption will provide a foundation for universally accepted design rules for high-performance SNSPDs and support the continued advancement of their record-setting metrics.

A number of models have been proposed to describe the response of SNSPDs after photon absorption; however, the mechanisms of superconductivity suppression and the dependence of the detection current on the photon impact position relative to the nanowire edges show varying levels of agreement with experimental data. This section discusses the main models that describe the operation of SNSPD.

\subsection{Electrical model}

SNSPD typically employs a superconducting nanowire with thickness ranging from 5 to 10~nm and width of approximately 100~nm, providing sensitivity of the superconducting state to low-energy single photons. The photon detection process in superconducting single-photon detectors can be described using electrothermal model, providing insight into device operation without requiring detailed analysis of microscopic processes\cite{yang2007modeling, kerman2009electrothermal}. 

\begin{figure}[htbp]
\centering\includegraphics[width=13cm]{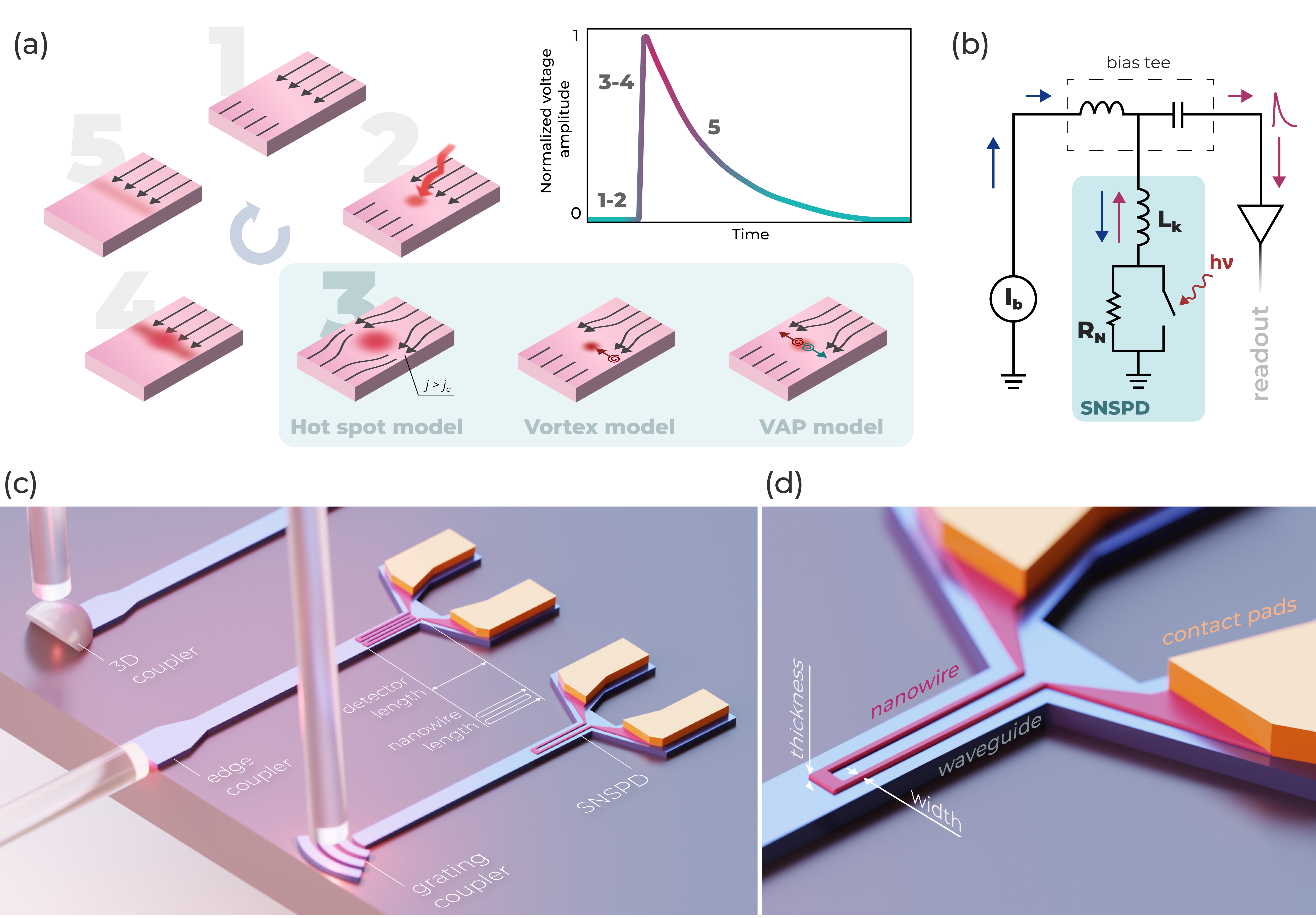}
\caption{(a) Simplified schematic of the SNSPD detection process. The inset illustrates a representative output pulse waveform of the detector, with distinct temporal regions annotated to correspond with key stages of the photon detection process. (b) Simplified bias and readout circuit of an SNSPD. (c) Overview of the conventional iSNSPD architecture. (d) Overview of the nanowire architecture.}
\label{fig:detection_models}
\end{figure}

A simplified schematic of the SNSPD detection process is shown in Fig.~\ref{fig:detection_models}(a) in five stages. For proper operation, the detector must be maintained at the temperature $T_{bath}$ well below the nanowire superconducting transition temperature $T_{c}$ (typically $T_{bath}$<4.2 K) and biased with a current $I_{b}$ that remains below the critical current of the nanowire $I_{c}$. This bias current serves two critical functions: it (1) preconditions the nanowire near the threshold of superconductivity breakdown, enabling photon-induced normal domain formation with minimal energy input, and (2) provides the driving force for measurable voltage pulse generation when the superconducting state becomes locally suppressed. When a photon is absorbed by the SNSPD, its energy locally disrupts the superconducting state, creating either a complete normal-conducting area or a region with a locally suppressed superconducting order parameter\cite{polyakova2022measuring}. This transition generates a measurable voltage pulse across the nanowire, which is amplified and measured by room-temperature readout electronics (e.g., oscilloscopes or time-correlated single-photon counting systems). The pulse front is determined by the duration of the resistive region formation stages, and the fall time is determined by the duration of superconductivity recovery. Superconductivity is restored through two primary cooling pathways: (1) thermal dissipation into the substrate, (2) quasiparticle recombination and heat diffusion along the nanowire\cite{yin2024heat}.

The electrical behavior of an SNSPD can be described by an equivalent circuit shown in Fig.~\ref{fig:detection_models}(b) consisting of a bias current source, readout electronics and detector itself, represented by: (1) a series kinetic inductance $L_{k}$, accounting for the inertia of superconducting carriers, (2) a parallel resistive branch modeling the normal-state resistance $R_{N}$ of the nanowire when triggered and (3) an ideal switch representing the transition between superconducting and resistive states\cite{kerman2009electrothermal}. This switch closes upon photon absorption, temporarily shunting current through $R_{N}$ and generating a measurable voltage pulse. 

Although the electrical model does not consider the process of superconductivity suppression caused by photon absorption in detail, it became the main focus of various theoretical models discussed below.  

\subsection{Hotspot model}

According to the hotspot model, photon absorption triggers the formation of a localized "hotspot" — a region with increased quasiparticle concentration (disrupted Cooper pairs)\cite{semenov2001quantum, semenov2002hot}. Between the hotspot and the nanowire boundaries, the local supercurrent density increases due to current flow around the resistive region. When the bias current is close to critical, local exceeding of the critical current density near the hotspot can lead to formation of a normal domain bridging the cross-section, resulting in a voltage pulse across the nanowire. The hotspot size evolution later has been associated with the diffusion of quasiparticles, taking into account that superconductivity in a hot spot can be suppressed, but not necessarily completely broken, and making the model more consistent with experimental results\cite{semenov2005spectral}.

Although this model has made it possible to successfully describe the detection process at high photon energies and bias currents, it implies that as either the bias current or photon energy decreases below a certain threshold, the detection efficiency should change abruptly from saturated value to zero\cite{semenov2005spectral}. However, experiments show that detection efficiency changes smoothly over a wide range of bias currents and photon energies\cite{cheng2019superconducting, wang2019wavelength}, which has confirmed the need to consider other detection models for a correct description of the SNSPD operation.

\subsection{Vortex model}

The vortex model proposes an alternative mechanism where photon absorption triggers magnetic vortices to entry the superconducting nanowire\cite{hofherr2010intrinsic, bulaevskii2011vortex}. In this framework, the absorbed photon locally suppress the superconducting order parameter, enabling vortex entry from the nanowire edges under the influence of the bias current. As these vortices crossing the nanowire width under the action of the Lorentz force, their motion induces a time-dependent voltage pulse\cite{jahani2020probabilistic}. According to some studies, the nanowire crossing by a single vortex is the dominant cause of detector dark counts\cite{yamashita2011origin, andreev2024dark}. 

Similar to the hotspot model, the vortex model has been supplemented by taking into account the diffusion of quasiparticles\cite{engel2013numerical, engel2014detection}, which has made it possible to describe experimental results related to the position of the absorbed photon. While this model successfully explains the gradual dependence of detection efficiency on bias current, it faces challenges in describing the dependence of the detection process on the width of the nanowire and bath temperature\cite{engel2015detection}.

\subsection{Vortex-antivortex model}

An alternative approach to photon detection in SNSPDs involves the generation and unbinding of vortex-antivortex pairs (VAP) in the superconducting nanowire \cite{semenov2008energy}. Unlike vortex models where single vortices enter the nanowire from edges, this mechanism considers the nucleation of bound vortex pairs within the superconductor volume following photon absorption \cite{wen2023improved}. 

When bias current flows through a superconducting strip, vortex-antivortex pairs align perpendicular to the current direction, but remain stationary, causing no energy dissipation and consequently generating no measurable voltage\cite{semenov2008vortex}. These bound pairs are stabilized by three competing forces: (1) the Lorentz forces driving vortices in opposite directions\cite{shi2025lorentz}, (2) their mutual magnetic attraction\cite{padavic2020vortex} and (3) pinning force, which arises due to the interaction of vortices with defects and inhomogeneities of the material\cite{jiang2022field}. Photon absorption disrupts this equilibrium by unbinding vortex-antivortex pairs, creating a resistive barrier across the strip\cite{engel2014detection}. Additionally, thermal fluctuations can cause spontaneous pair dissociation, contributing to dark count events\cite{yamashita2011origin}.

Some models consider the role in the detection process of both vortices entering the strip from the edge and the unbinding of vortex–antivortex pairs \cite{zotova2012photon, zotova2014intrinsic, vodolazov2014current, vodolazov2015vortex}. This approach allows achieving a high degree of agreement with experimental results, such as the width and temperature dependencies. However, it still does not explain certain experimental results related to position dependence.

\subsection{Summary}

Modern understanding recognizes that no single model fully explains all experimental observations across different device geometries and operating conditions. Often, modern studies consider a hybrid approach where hotspot formation dominates in narrow nanowires at high photon energies (or high bias currents), and single vortex or VAP dynamics become significant in wider nanowires and lower photon energies \cite{raj2025waveguide}. Nanowire geometric and material properties, along with operating conditions, determine the relative contribution of each mechanism. Although the development of a universal model is still a challenge, the experimental dependencies of various SNSPD performance metrics on its parameters are well studied and will be discussed further.

\section{iSNSPD performance metrics}

Different types of single-photon detectors are characterized by various performance metrics due to the specifics of their operation and target applications\cite{wang2025advances}. This consideration is also relevant when comparing stand-alone SNSPDs with integrated devices: certain performance metrics differ, whereas others, although identical in value, are governed by entirely different mechanisms. In this section, we review the key parameters used to evaluate iSNSPDs and analyze how different detector performance metrics depend on nanowire geometry, material properties, operating conditions and light properties.

The conventional integrated SNSPD configuration consists of three main elements: an optical input coupler, the waveguide itself and a superconducting nanowire with contact pads monolithically integrated directly on the waveguide as shown in Fig.~\ref{fig:detection_models}(c). Several implementations integrate an inductive element, connected in series with the iSNSPD, to mitigate latching effects and ensure stable operation under high photon flux conditions \cite{sahin2013waveguide}. Alternative designs and integration approaches will be discussed in Section 6.

During device design and fabrication, particular attention is given to detector parameters that determine overall performance, including geometric factors and the properties of the superconducting material\cite{henrich2014influence}. Unlike conventional meander-patterned stand-alone detectors, integrated SNSPDs typically use either a U-shaped single-loop or W-shaped double-loop nanowire configuration. The W-shape provides a larger interaction area with the waveguide while maintaining the same device footprint. It is crucial to distinguish between the total SNSPD length, which determines the device footprint and consequently the technology scalability, and the nanowire length, meaning the length of the U-shape or W-shape curve that governs the detector response dynamics. Although different references may adopt varying terminology, this review will consistently use the terms "detector length" and "nanowire length", respectively. The geometric parameters of the nanowire that determine the iSNSPD sensitivity to single photons are its width and thickness (see Fig.~\ref{fig:detection_models}(d)), since they determine the photon energy required to suppress or break superconductivity across the entire cross-section \cite{henrich2014influence}.

Beyond the detector intrinsic parameters that can be chosen during design and fabrication, its characteristics are also influenced by the performance of the photonic integrated circuit (PIC), device operating conditions and light properties \cite{ferrari2019characterization}. The impact of each of the mentioned above parameters on iSNSPD performance will be discussed in detail in this section.

\begin{figure}[htbp]
\centering\includegraphics[width=13.5cm]{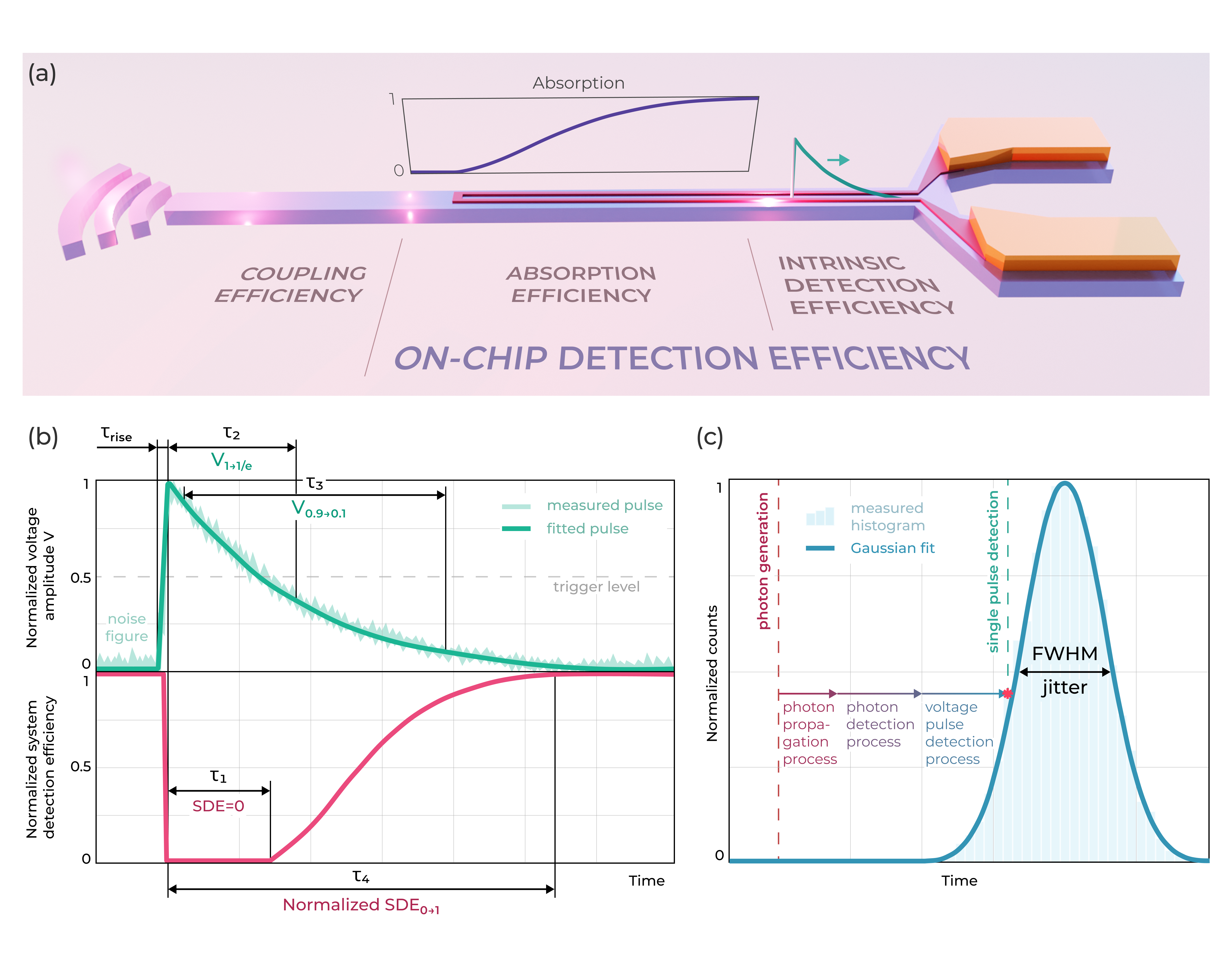}
\caption{Integrated superconducting single-photon detector performance overview. (a) Visualization of the components of the system detection efficiency of the iSNSPD. (b) Correlation between the detector output pulse shape and the time-dependent detection efficiency recovery. (c) Statistical distribution of photon arrival times, illustrating timing jitter determination.}
\label{fig:metrics}
\end{figure}

\subsection{System detection efficiency}

The primary characteristic of single-photon detectors is their system detection efficiency (SDE), which represents the probability that a photon entering the system will generate a registered response. The "system" here includes both the measurement circuitry and the cryostat with optical interfaces, fiber-to-detector coupling, the detector itself and the readout circuit \cite{henrich2014influence}. For iSNSPDs, SDE can be decomposed into coupling, absorption and intrinsic detection efficiency, as shown in Fig.~\ref{fig:metrics}(a). The contribution of each factor will be discussed below.

\subsection{Coupling efficiency}

The coupling efficiency represents the probability that a photon entering the system will reach the SNSPD active area. This component accounts for propagation losses in optical fibers and waveguides, but primarily depends on the alignment between the optical fiber and the waveguide or detector \cite{baker2018superconducting}. In integrated source-detector systems, the efficiency of coupling light from the on-chip source into the waveguide should be also treated as a main component of the coupling efficiency.  

For iSNSPDs coupling efficiency depends mainly on the input coupling technique and on-chip losses, though the latter is typically negligible compared to the former\cite{wolff2021broadband,ferrari2019characterization}. Common waveguide coupling approaches include grating couplers\cite{lomonte2024scalable}, edge couplers\cite{yi2025polarization} and emerging 3D photonic structures\cite{dietrich2018situ}. Moreover, the efficiency of a coupling element depends not only on its design and fabrication quality but is also critically influenced by external operational parameters, including operating temperature, incident light polarization and photon wavelength \cite{cheng2020grating,fahrenkopf2019aim}. Thus, the coupling methodology itself is equally important due to cryogenic operation challenges.

\subsection{Absorption efficiency}

Absorption efficiency corresponds to the probability that a photon reaching the active area will be absorbed by the nanowire. This metric depends on the nanowire material absorption coefficient and its geometry\cite{semenov2009optical,hydyrova2021evaluation}, as well as on light polarization\cite{guo2015single,miki2010characterization,ferrari2019characterization}.

For iSNSPDs absorption efficiency is primarily governed by the nanowire-light interaction area determined by nanowire length\cite{mattioli2020waveguide,wolff2020superconducting}, width\cite{kahl2015waveguide,baranek2024numerical,li2019design} and thickness\cite{baranek2024numerical, strohauer2023site}. For example, W-shaped nanowires can increase absorption without enlarging device footprint\cite{kovalyuk2017chip, majety2023triangular}, however, longer nanowires not only increase device size but also raise defect probability during fabrication\cite{constancias2007patterning}, increase dark count rate and reduce maximum count rate\cite{wolff2020superconducting}. Alternative enhancement methods include integration of SNSPD into photonic crystal cavities in waveguides \cite{akhlaghi2015waveguide}, as will be discussed in Section 6.

\subsection{Intrinsic detection efficiency}

Intrinsic (or internal) detection efficiency (IDE) represents the probability that an absorbed photon will generate a detector response. This SNSPD-specific metric characterizes its sensitivity to particular photon wavelengths\cite{wang2019wavelength}. IDE has complex dependencies on numerous parameters common to both stand-alone and integrated SNSPDs, primarily nanowire cross-section\cite{henrich2014influence,azem2024mid} and superconducting material properties\cite{henrich2014influence, zolotov2021comparison, strohauer2023site, tripathy2024comparative}. While reducing nanowire width and thickness can significantly improve IDE, technological limits (particularly electron beam lithography resolution limit) \cite{marsili2011single} and dark counts\cite{martini2020electro} impose practical limits. Current IDE optimization therefore focuses mainly on material engineering\cite{tripathy2024comparative,haldar2025influence,strohauer2025origin}, though the exact influence of superconducting properties is still under discussion due to complex interparameter dependencies\cite{engel2013temperature,stepanov2024sputtered}. The IDE shows strong dependence on both bias current and wavelength. With decreasing wavelength the higher photon energy enhances superconductivity suppression, increasing the probability of detector pulse generation, while at a fixed wavelength, higher bias currents improve IDE by reducing the required energy for local weakening of the superconducting state needed for photon detection\cite{zhang2019saturating, wang2019wavelength}. The IDE metric is also closely related to the SNSPD count rate, since photons arriving during the detection process cannot be registered. Thus, after a photon is detected, its intrinsic detection efficiency remains zero for a certain time, and then increases to the nominal value\cite{autebert2020direct}, so careful calibration of the photon flux is necessary to obtain the highest efficiency\cite{chang2021detecting}.

\subsection{Threshold efficiency}

The threshold efficiency or measurement system efficiency, often assumed to be unity, describes the probability that a detector pulse caused by the photon response will be properly registered by readout electronics. For conventional SNSPD readout it depends on the voltage pulse amplitude-to-noise ratio and trigger threshold settings. For optimal operation, the noise amplitude should lie significantly below the trigger threshold, with SNSPD pulses exceeding it by a factor of two\cite{ferrari2018waveguide}. The voltage pulse amplitude $U$ itself depends on the bias current $I_{b}$, which in turn is limited by the critical current $I_{c}$. Nanowire critical current is determined by the nanowire geometry\cite{baghdadi2021enhancing, lita2015materials} and critical current density of superconducting material\cite{yang2017comparison, xiong2022reducing}, which is influenced by bath temperature and input photon flux\cite{yabuno2023superconducting,kovalyuk2017chip}.The SNSPD pulse amplitude also depends on signal losses due to impedance mismatches between the detectors and the readout circuit, which is why some papers demonstrate the use of impedance tapers, which not only increase the pulse amplitude several times, but also reveal the detector ability to resolve the number of photons\cite{colangelo2023impedance, zhu2019superconducting, zhu2020resolving}, as will be discussed in Section 5.

\subsection{On-chip detection efficiency}

On-chip detection efficiency (OCDE) serves as the primary metric for integrated SNSPDs. Unlike system detection efficiency, OCDE considers only absorption and intrinsic detection efficiencies, excluding coupling losses. This metric characterizes the detector independently from photonic integrated circuit parameters, particularly input coupling losses, which are a separate research focus\cite{avdeev2024tutorial}. Most studies report OCDE for iSNSPDs, while SDE is less frequently used since achieving high coupling efficiency is not the primary goal of most articles demonstrating on-chip detector integration. Some studies use detection efficiency (DE) for standalone SNSPDs - conceptually equivalent to OCDE, although experimentally separating coupling losses remains technically challenging\cite{dauler2014review}.

\subsection{Dark count rate}

The dark count rate (DCR) represents the number of false detector responses per second. These anomalous counts arise primarily from geometric effects, particularly current crowding phenomena at nanowire bends and fundamental material limitations \cite{clem2011geometry, xiong2022reducing}. Recent advances in stand-alone SNSPD designs have demonstrated that optimized nanowire geometries\cite{stepanov2022superconducting, reddy2022broadband, jonsson2022current} and local material engineering\cite{strohauer2025current} can substantially mitigate current crowding effect. Beyond geometric factors, DCR is influenced by superconducting material fluctuations\cite{engel2015detection,xu2021superconducting} which can be reduced by lowering the operating temperature and bias current\cite{shehata2015effect,zhang2014characterization}, blackbody radiation \cite{shibata2013superconducting} and external noise sources\cite{natarajan2012superconducting,chen2015dark}, the impact of which can be minimized using cryogenic shielding \cite{10.1063/5.0250262}. Current research continues to actively investigate the fundamental origins of dark counts\cite{zhang2022geometric}. Due to their shorter nanowire length, fewer geometric bends and absence of open-space coupling, iSNSPDs typically exhibit lower dark count rates compared to their stand-alone counterparts.

\subsection{Count rate and recovery time}

The maximum count rate (CR) represents the highest number of photons per second that an SNSPD can detect. While this metric is crucial for practical device implementation, the existing research typically reports the count-rate-determining recovery time $\tau _{recovery}$\cite{wang2023controlling,oshiro2025accurately,he2015statistical} (also referred to as reset time\cite{annunziata2010reset,zhao2014counting,dong2024establishing} or decay time\cite{verma2015high,verma2014superconducting,wollman2017uv} in some publications). This temporal characteristic admits multiple interpretations across different research groups. Although the relationship $CR=1/\tau _{recovery}$ is commonly cited\cite{kerman2006kinetic}, the definition of recovery time itself varies substantially. Fig.~\ref{fig:metrics}(b) illustrates the various time metrics commonly employed to describe SNSPD response dynamics. Some studies define recovery time as the duration required for the detector pulse amplitude to decay by a factor of $e$ ($\tau_{2}$ or 1/$e$ recovery time)\cite{chang2021detecting,verma2014superconducting,tripathy2024comparative,jiang2024superconducting} or to fall from 90\% to 10\% of its peak value ($\tau_{3}$)\cite{wollman2017uv,gu2015high,vyhnalek2018performance}, while others characterize it as the time needed for detection efficiency to return to its nominal value ($\tau_{4}$)\cite{autebert2020direct,he2015statistical}. Furthermore, the term dead time appears in literature with divergent meanings - sometimes used synonymously with recovery time\cite{dong2024establishing}, while in other contexts denoting the absolute refractory period when detector cannot register subsequent photons ($\tau_{1}$)\cite{uzunova2022photocounting,ferrari2018waveguide}. Thus, this metric not only suffers from inconsistent interpretation across studies, but also lacks standardized measurement methodologies. 

Recovery time is fundamentally governed by the properties of incoming light\cite{chang2021detecting, autebert2020direct} and the nanowire kinetic inductance $L_{k}$, which depends on the nanowire geometry and material properties\cite{kerman2006kinetic,yang2017comparison,tripathy2024comparative}. While CR can be enhanced by reducing nanowire length and optimizing its cross-sectional dimensions\cite{henrich2014influence}, such improvements face a critical limitation: latching, a failure mode where the detector persists in a resistive state post-detection, unable to reset to superconductivity\cite{liu2012nonlatching}. Therefore, careful engineering of the SNSPD parameters is necessary for its reliable dynamic operation. 

\subsection{Jitter}

Timing jitter in SNSPDs refers to the temporal uncertainty between the photon generation and the actual registered detection event, representing a fundamental resolution limit for time-correlated applications. Experimentally, jitter is quantified as the full width at half maximum (FWHM) of the histogram of time delays between sending a photon into the system (in practice, synchronized laser pulses) and the SNSPD responses (Fig.~\ref{fig:metrics}(c)). The measured detector jitter is conventionally divided into several components, each with distinct origins. Optical jitter describes the variability in time between photon generation and its arrival at the detector, determined primarily by fiber dispersion\cite{ferrari2019characterization}. The detector intrinsic jitter represents the uncertainty in the duration of the detection process itself, including hotspot formation time and voltage pulse propagation \cite{allmaras2019intrinsic,caloz2018high}. This metric depends on material properties \cite{yang2017comparison}, operation conditions (such as bias current and bath temperature) affecting the dynamics of hot spot growth, and nanowire length governing signal propagation time\cite{gourgues2019superconducting, calandri2016superconducting,santavicca2019jitter}. Electrical jitter shows variations in pulse registration timing, being highly dependent on noise introduced by the readout circuitry and the set trigger threshold level\cite{you2013jitter,ferrari2019characterization}.

\subsection{Summary}

\begin{figure}[htbp]
\centering\includegraphics[width=13.5cm]{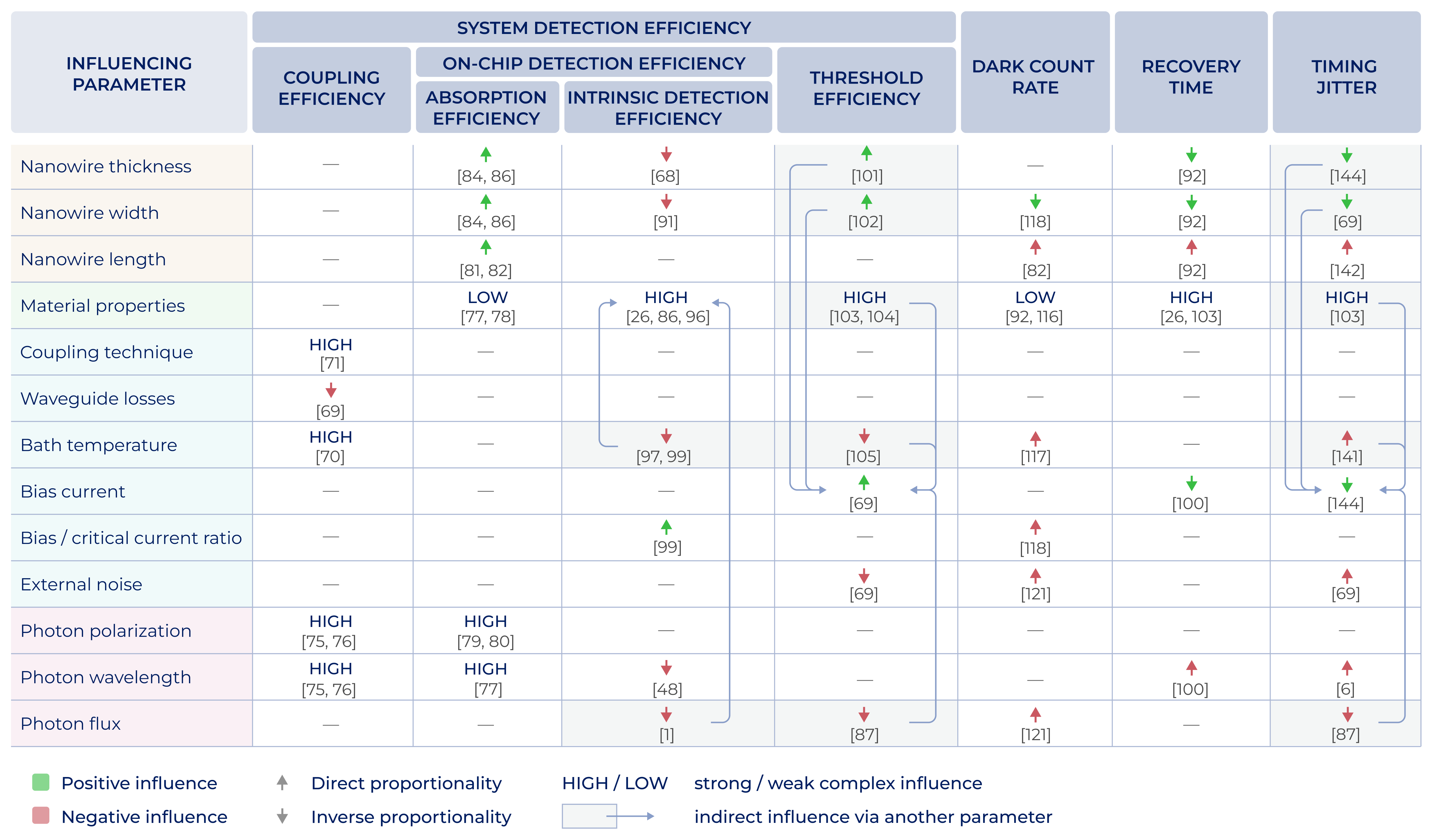}
\caption{Overview of the dependence of iSNSPD performance metrics on detector parameters, operational environment and incident photon properties.}
\label{fig:metrics2}
\end{figure}

Fig.~\ref{fig:metrics2} summarizes how integrated SNSPD parameters (superconducting material, nanowire geometry, operating conditions and light properties) affect the key detector metrics discussed in this section. This comprehensive parameter space analysis establishes the foundation for systematic iSNSPD optimization, guiding both device engineers in achieving target specifications and application specialists in selecting appropriate operating regimes for their specific use cases. 

Further improvement of the discussed performance metrics and their reproducible fabrication at industrial scales represents a critical step for practical iSNSPD applications. To achieve this, a systematic understanding of the interdependencies between these characteristics and various device parameters is essential, including both design-controlled factors and application-specific constraints. Such comprehensive optimization requires coordinated advances in three domains: (1) fundamental physics of photon-induced superconductivity suppression to overcome trade-offs of performance metrics and the possibility of designing a detector with characteristics satisfying the application, (2) nanofabrication techniques ensuring the achievement of the designed detector parameters and its uniformity across large-scale arrays, (3) standardized and optimized (in terms of both impact on metrics and price) techniques for detector readout and maintaining suitable operating conditions.

\section{Detector and waveguide materials}

As shown in the previous section, the properties of the superconducting material affect almost all iSNSPD performance metrics. At the same time, the detector integrated into the waveguide is a step towards realizing scalable devices that combine multiple active and passive components on a single chip. The integrated photonics platform, although it does not directly affect the performance of the detector, is crucial for creating iSNSPD-based devices and determines their scalability and capabilities. This section reviews the main superconducting and waveguide materials used to fabricate integrated SNSPDs, a comprehensive analysis of detector performance achieved across diverse material platforms and their development prospects.

\subsection{Superconducting materials for iSNSPDs}

The superconducting materials used in integrated SNSPDs can be divided into two groups: crystalline (nitrides), including niobium nitride (NbN) and niobium titanium nitride (NbTiN), and amorphous (silicides), including tungsten silicide (WSi) and molybdenum silicide (MoSi). Promising superconductors whose waveguide integration has not yet been demonstrated are also discussed in this section.

\subsubsection{Niobium nitride (NbN)}

Niobium nitride has been the first material to demonstrate single-photon detection of SNSPD in both stand-alone\cite{gol2001picosecond} and waveguide-integrated implementations\cite{sprengers2011waveguide}. Traditionally used in superconducting bolometers\cite{weiser1981use}, NbN has become the historically dominant material for SNSPDs due to both well-studied deposition methods\cite{akune1982nbn} and its outstanding properties, foremost a relatively high critical temperature (bulk $T_{c} \approx$ 16 K\cite{gavaler1971superconducting}) allowing the use of commercially available two-stage helium-4 cryocoolers and at the same time achieve high sensitivity. An important material parameter is the relaxation time $\tau_{th}$, which characterizes the process of energy dissipation after the photon detection and determines the detector response dynamics. Due to the short relaxation time on the order of tens of ps\cite{zhang2018hotspot} and high $T_{c}$ NbN remains actively investigated and widely used material for single-photon detectors\cite{stepanov2024sputtered,zhang202332,zhang2024sputtering}. For iSNSPDs, NbN has enabled over 90\% on-chip detection efficiency at telecom wavelengths combined with outstanding dynamic performance, including sub-20-ps timing jitter and sub-nanosecond recovery time\cite{pernice2012high}. Moreover, this material has been chosen by the PsiQuantum company for integrated photonic quantum computing platform, demonstrating near-unity on-chip detection efficiency (median 98.9\% and mean 96.2$\pm$4.3\%)\cite{psiquantum2025manufacturable}. NbN-based detectors hold records for dark count rate of 0.0001 Hz\cite{munzberg2018superconducting} and recovery time of 120 ps\cite{vetter2016cavity}, both made possible by reducing the length of the nanowire to a few micrometers. However, in order to improve the detector sensitivity to near-infrared photons and response dynamics, other superconducting materials have been studied.

\subsubsection{Niobium titanium nitride (NbTiN)}

Despite the fact that NbN and NbTiN have similar electronic and superconducting properties, the latter has increasingly attracted the attention of researchers in recent years due to a number of its advantageous differences\cite{dong2024establishing, azem2024mid, china2023highly}. Niobium titanium nitride offers reduced resistivity, lower kinetic inductance and higher critical current density compared to NbN\cite{yang2017comparison,miki2009superconducting}, combined with even slightly higher critical temperature (bulk $T_{c}$ up to 18 K\cite{peeters2025superconducting}). These material properties allow for significant improvements in recovery time and jitter with comparable sensitivity to single photons. Thus, using NbTiN, one of the lowest jitter for iSNSPDs published to date equal to 12.2 ps has been demonstrated, along with an on-chip detection efficiency of 88\%\cite{beutel2022cryo, schuck2024integration}. Moreover, along with niobium nitride, NbTiN-based detector has demonstrated near-unity OCDE combined with a DCR of about 0.1 Hz, timing jitter of 55 ps and recovery time of 7 ns using a nanowire only 8.5 $\mu$m long, which has been achieved by integrating the detector into photonic crystal cavity\cite{akhlaghi2015waveguide}. 

Although the use of NbN and NbTiN allows combining high efficiency and fast response dynamics, for telecommunication wavelengths, and all the more so for the mid-IR range, it is still difficult to achieve a saturated intrinsic detection efficiency due to the relatively large energy gap. One approach to increase the sensitivity of crystalline superconductor based detectors to longer photon wavelengths is to tune the material properties by ion irradiation. The improvement in SNSPD performance by helium ion irradiation has been first demonstrated for NbN\cite{zhang2019saturating, hong2025impact} and then for NbTiN\cite{strohauer2023site, wietschorke2025enhancement}. This effect has initially been associated with a irradiation-induced vacancies, increasing film disorder, decreasing the superconducting energy gap and the electron density of states at the Fermi level\cite{zhang2019saturating}. Strohauer et al. have conducted a comprehensive study on the origins for the improved performance of ion-irradiated SNSPDs, showing that in addition to the defect engineering in NbTiN film, a significant contribution has come from the reduction in thermal conductivity due to substrate amorphization and a modified superconductor-substrate interface, increasing the intrinsic detection efficiency\cite{strohauer2025origin}. This approach to increase the sensitivity has not yet been applied to iSNSPDs and has significant technological limitations, since it would affect the properties of the waveguide material and the waveguide-superconductor interface. Nevertheless, precise selection of the dose and energy of irradiation and design of systems with pre-accounted changes in their properties can make this technology extremely promising for achieving outstanding performance.

The relatively large energy gap is not the only drawback of crystalline superconductors. The substrate choice and the crystal structure influence the superconducting properties of the material, which can also pose significant limitations for waveguide integration; furthermore, inhomogeneities of the crystalline phase reduce the yield \cite{toth2014transition}. Based on this, amorphous superconducting materials with a significantly smaller energy gap have been proposed.

\subsubsection{Tungsten silicide (WSi)}

Tungsten silicide belongs to a class of amorphous superconductors that exhibit enhanced uniformity over large areas\cite{zhang2016superconducting}, an advantage for large-scale devices with reproducible performance and detector arrays\cite{wollman2019kilopixel}. Thus, WSi has been chosen by Oripov et al. to demonstrate an array of stand-alone SNSPDs of more than 400,000 pixels\cite{oripov2023superconducting}. Although this material has a higher kinetic inductance than crystalline superconductors\cite{hao2024compact} and requires less cost-effective cryostats to provide <1 K bath temperature for operation due to WSi low critical temperature (bulk $T_{c} \approx$ 5 K\cite{verma2014superconducting}), it also attracts the attention of researchers because of its small energy gap and, as a result, high sensitivity to photons of mid-IR range\cite{marsili2013detecting,chen2018ultra,verma2021single}. A few WSi-based iSNSPDs have been demonstrated to date. McDonald et al. have demonstrated the integration of a WSi-based detector with a light-emitting diode on a III-V platform, claiming outstanding DCR of less than $10^{-3}$ Hz and saturated intrinsic detection efficiency at 895 nm, with the OCDE estimated as >90\% based on absorption efficiency simulation \cite{mcdonald2019iii}. The only paper reporting measured detection probability has demonstrated a system detection efficiency of 2.5\% together with a saturated IDE at 1615 nm, which represents a competitive result; however, OCDE has not been reported \cite{beyer2015waveguide}. Nevertheless, the disadvantages of amorphous superconductors are longer recovery times and jitter, which for waveguide-integrated WSi-based SNSPDs have been demonstrated to be only 30 ns\cite{shainline2017room} and 380 ps\cite{hopker2021integrated}, respectively. Despite the small number of references to tungsten silicide in the context of iSNSPD, this material is noted by researchers as one of the most promising, especially for the realization of integrated detectors arrays for scalable quantum information processing systems\cite{buckley2017all, hopker2021integrated}.

\subsubsection{Molybdenium silicide (MoSi)}

Molybdenium silicide is another amorphous superconductor known for the near-unity yield provided even when deposited at room temperature\cite{erbe2024mo}. With an even slightly higher critical temperature than WSi (bulk $T_{c} \approx$ 7.5 K\cite{banerjee2017characterisation}), and thus allowing comparable detection efficiency to be achieved even at higher temperatures, MoSi is a promising material for exploiting the advantages of an amorphous superconductor without the need for cooling to <1 K. Recent MoSi-iSNSPD implementations have achieved on-chip detection efficiencies up to 73\% combined with sub-5 ns recovery time, comparable to those of crystalline materials, and DCR of 10 Hz, though accompanied by a relatively high jitter of 135 ps\cite{haussler2020amorphous}. Thus, high jitter still remains one of the main limitations of molybdenium silicide application. A minimum jitter of 51 ps had been achieved on MoSi-based iSNSPD \cite{li2016nano}. As with WSi, the number of papers on MoSi-iSNSPDs is smaller than for crystalline materials. Nevertheless, this material is attracting increasing attention, and MoSi-based devices are being actively explored \cite{grotowski2025optimizing, colangelo2024molybdenum}.

Table \ref{tab:supercond} compares the properties of superconducting materials that have been used for integrated single-photon detectors. The parameters chosen for comparison are: (1) the bulk critical temperature, since it determines the $T_{c}$ of the ultrathin film and, as a consequence, the required operating temperature, (2) the critical current density $J_{c}$, which affects jitter and DCR, (3) the zero-temperature magnetic penetration depth $\lambda(0)$, the square of which is inversely proportional to the recovery time\cite{engel2012tantalum}, as well as (4) the material structure, which determines its features discussed in this section. It is important to note that since all of these materials are two-component, the properties presented may vary significantly depending on the stoichiometry of the film and its deposition mode.

\smallskip

\renewcommand\tabularxcolumn[1]{m{#1}}
\begin{table}[htbp]
\caption{Comparison of superconducting materials properties}
  \label{tab:supercond}
  \centering
\begin{tabularx}{1\textwidth}{ >{\centering\arraybackslash}X >{\centering\arraybackslash}X >{\centering\arraybackslash}X >{\centering\arraybackslash}X >{\centering\arraybackslash}X}
\hline
Property & NbN & NbTiN & WSi & MoSi \\
\hline
Bulk $T_{c}$, K & 16 \cite{gavaler1971superconducting} & 18 \cite{peeters2025superconducting} & 5 \cite{verma2014superconducting} & 7.5 \cite{banerjee2017characterisation}\\
Critical current density $J_{c}$, $MA/cm^{2}$ & up to 17 @4.2K\cite{charaev2017enhancement} & up to 19 @2K \cite{zhang2024effect} & up to 1 @0.25K\cite{lita2015materials}  & up to 2.5 @0.25K\cite{korneeva2014superconducting}\\
Magnetic penetration depth for thin film $\lambda(0)$, nm & 400 \cite{bartolf2010current} & N/A & 680-770 \cite{zhang2016characteristics} & 500-730 \cite{zhang2021physical} \\
Structure & Crystalline & Crystalline & Amorphous & Amorphous \\
\hline
\end{tabularx}
\end{table}

\begin{figure}[htbp]
\centering\includegraphics[width=13.5cm]{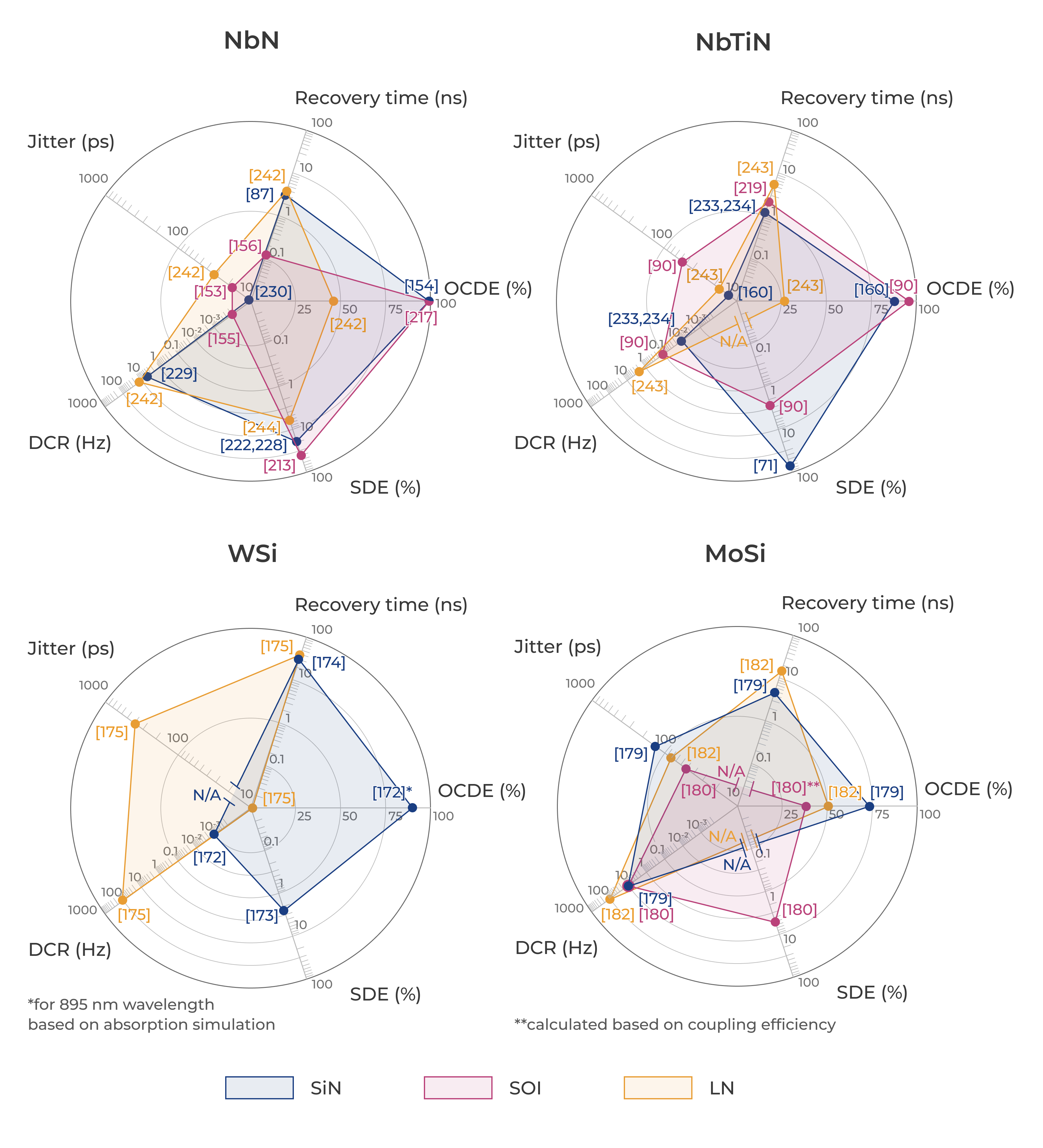}
\caption{Record performance metrics of superconducting single-photon detectors based on various superconducting materials integrated on different photonic platforms.}
\label{fig:radars}
\end{figure}

Fig.~\ref{fig:radars} illustrates the record performance metrics achieved by iSNSPDs based on various superconducting materials and photonic platforms, providing a clear comparison of their benefits and limitations in the context of future applications.

Thus, materials considered demonstrate the most promising performance indicators and remain strong candidates for further technological development not only in stand-alone, but also in waveguide-integrated detectors\cite{tripathy2024comparative}.

\subsubsection{Promising superconducting materials}

Beyond the established materials, several alternatives are under active investigation to overcome existing trade-offs. None of the materials discussed below have been demonstrated in a waveguide integrated implementation, but since they have been used for stand-alone SNSPDs and have shown some encouraging results, we highlight them as promising for iSNSPDs as well.

Using crystalline niobium-based compounds NbC\cite{korneeva2016comparison} and NbSi\cite{dorenbos2011low}, detection of single photons with wavelengths of 405 and 1900 nm, respectively, has been demonstrated, but due to the limited sensitivity of these materials, studies of detectors based on them have not been encountered in recent years. In contrast, niobium rhenium (NbRe) is an actively studied material for use in single-photon detectors to improve sensitivity without sacrificing timing performance\cite{caputo2017nbre}. NbRe-based SNSPD has recently demonstrated saturated IDE at 1300 nm with recovery time below 10 ns and jitter of about 28 ps, which is comparable to typical metrics of NbN and NbTiN-based detectors \cite{avitabile2025superconducting}. In addition, NbRe has been used to demonstrate detection of single photons at telecommunication wavelengths using a 2.2$\mu$m wide strip\cite{ejrnaes2022single,cirillo2024single}, which confirms its potential for the development of SMSPDs (superconducting microstrip single-photon detectors), and research into this material and the possibilities of its modification continues\cite{cirillo2021nbren}.

Amorphous materials are also being actively studied for use in single-photon detectors. Thus, detectors based on amorphous tungsten germanide (WGe) with superconducting properties similar to those of tungsten silicide have demonstrated jitter as low as 127 ps\cite{yang2025superconducting} and saturated IDE for photons with wavelengths up to 29 $\mu$m \cite{hampel2025tungsten}, highlighting exceptional sensitivity and promising potential of this material. Among the amorphous Mo-based superconducting materials that are still being studied today for use in SNSPD molybdenum nitride (MoN) and molybdenum–rhenium (MoRe) can be distinguished. Molybdenum nitride, which has a critical temperature comparable to MoSi, has made it possible to achieve saturated intrinsic detection efficiency for visible\cite{nishikawa2022fabrication} and telecom\cite{hallett2021superconducting} photons, while MoRe-based detectors have demonstrated IDE of 98\% and 73.5\% for wavelengths of 780 and 1550 nm, respectively \cite{nevzorov2025more}.

High-temperature superconducting materials represent a highly promising direction for advancing SNSPDs. Recent research highlights that these materials enable single-photon detection at temperatures significantly higher than those required by conventional low-temperature superconductors. Charaev et al.\cite{charaev2023single} have marked a significant breakthrough in the development of SNSPDs based on high-temperature superconductors, demonstrating for the first time single-photon detection in nanowires fabricated from $Bi_{2}SR_{2}CaCu_{2}O_{8+\delta}$ (BSCCO) thin flakes and $La_{1.55}Sr_{0.45}CuO_{4}/La_{2}CuO_{4}$ (LSCO–LCO) bilayer films at temperatures of 25 and 8 K, respectively. A key technological advance has been the use of a focused ion beam for: (1) patterning structures from BSCCO flakes, which are incompatible with standard nanofabrication processes and (2) inducing the necessary current-voltage hysteresis for photon detection in LSCO–LCO devices via helium ion irradiation. In addition to these high-temperature superconductors, magnesium diboride (MgB$_{2}$), with its bulk critical temperature of about 39 K\cite{nagamatsu2001superconductivity} has demonstrated robust single-photon detection using a micrometer wide strip at temperatures up to 20K and saturated IDE for a wavelength of 1.5 $\mu$m at a temperature of 3.7 K\cite{charaev2024single}. This revolutionary result has been also made possible by using helium ion irradiation process that affects both the MgB$_{2}$ film properties and the defects and amorphization in the substrate. Recent advances in the use of high-temperature superconductors for single-photon detection demonstrate the connection between material engineering not only with the possibility of increasing the SNSPD sensitivity, but also with an increase in its operating temperature and wire width, which determine the cost of the system due to the required cryogenic technology and equipment for nanofabrication, respectively. 

Table~\ref{tab:supercond_prom} summarizes the achieved performance of standalone SNSPDs based on major promising superconducting materials, highlighting their potential sensitivity and timing performance.

\newpage

\renewcommand\tabularxcolumn[1]{m{#1}}
\begin{table}[htbp]
\caption{Performance of standalone SNSPDs based on promising superconducting materials}
  \label{tab:supercond_prom}
  \centering
\begin{tabularx}{\textwidth}{ 
        >{\raggedright\arraybackslash}m{2cm} 
        >{\centering\arraybackslash}m{1.5cm} 
        >{\centering\arraybackslash}m{1.5cm} 
        >{\centering\arraybackslash}m{1.5cm} 
        >{\centering\arraybackslash}m{1.5cm}
        >{\centering\arraybackslash}m{3cm}
    }
\hline
Metric & NbRe & WGe & MoN & MoRe & MgB$_{2}$ \\
\hline
Max. IDE & \makecell{100\% \\ @1.3 $\mu$m \\ \cite{avitabile2025superconducting}} & \makecell{100\% \\ @29 $\mu$m \\ \cite{hampel2025tungsten}} & \makecell{100\% \\ @1.5 $\mu$m \\ \cite{hallett2021superconducting}} & \makecell{98\% \\ @0.78 $\mu$m \\ \cite{nevzorov2025more}} & \makecell{100\% (using $\mu$m \\ wide strip) @1.5 $\mu$m \\ \cite{charaev2024single}} \\
Min. jitter, ps & 28 \cite{avitabile2025superconducting} & 127 \cite{yang2025superconducting} & 50 \cite{hallett2021superconducting} & 132 \cite{nevzorov2025more} & 50 \cite{charaev2024single} \\
Min. recovery time, ns & 8 \cite{avitabile2025superconducting} & — & — & 4.1 \cite{nevzorov2025more} & 1.3 \cite{charaev2024single} \\
\hline
\end{tabularx}
\end{table}

Thus, although the materials considered have not yet been used in integrated detectors, they may be promising for further improving the performance of iSNSPDs for integrated quantum systems, both in terms of enhancing sensitivity and increasing the operating temperature.

\subsection{Waveguide materials for iSNSPDs}

The photonic platform underlying the iSNSPD is required to support low-loss guiding at target wavelengths, efficient light coupling and compatibility with scalable fabrication. Furthermore, modern trends highlight the need to integrate additional active components, including single-photon and coherent sources as well as high-speed modulators \cite{lu2024emerging, xu2023mid, aghaee2025scaling}. Waveguide materials determine the availability of active functionality and compatibility with hybrid integration technologies\cite{chen2024hybrid} and thus play a foundational role in systems with integrated SNSPDs. 

Table \ref{tab:overview} summarizes reported demonstrations of integrated SNSPDs across different photonic platforms and superconducting materials. This overview highlights the range of superconductor-waveguide materials combinations that have been experimentally realized and provides insight into current trends in the field. 

\begin{table}[htbp]
\caption{Overview of the iSNSPDs based on various superconductors integrated on different photonic platforms and their record efficiency metrics}
  \label{tab:overview}
  \centering
\begin{tabularx}{\textwidth}{ 
        >{\raggedright\arraybackslash}m{1.9cm} 
        >{\centering\arraybackslash}m{1.9cm} 
        >{\centering\arraybackslash}m{1.8cm} 
        >{\centering\arraybackslash}m{1.7cm} 
        >{\centering\arraybackslash}m{1.7cm} |
        >{\centering\arraybackslash}m{1.8cm}
    }
\hline
Photonic platform & NbN & NbTiN & WSi & MoSi & Max. SDE \\
\hline
SOI & \cite{pernice2012high, schuck2013matrix, vetter2016cavity, munzberg2018superconducting, shibata2019waveguide, yu2019silicon, yun2020superconducting, ono2021si, zheng2021heterogeneously, zheng2023chip, shu2025chip, psiquantum2025manufacturable, li2025surpassing} & \cite{akhlaghi2015waveguide, azem2024mid, tao2025single, martinelli2026inline} & \cite{buckley2017all, buckley2020integrated} & \cite{li2016nano} & 40\% \cite{ono2021si} 32\% \cite{shibata2019waveguide} \\
\hline
SiN & \cite{cavalier2011light, najafi2015chip, kahl2015waveguide, khasminskaya2016fully, kovalyuk2017chip, kahl2017spectrally, gaggero2019amplitude, elsinger2019integration, hartmann2020broadband, beutel2022fully, cheng2023100, schutte2023waveguide, psiquantum2025manufacturable, page2025scalable} & \cite{schuck2013optical, schuck2013waveguide, schuck2013nbtin, schuck2016quantum, guo2017parametric, gourgues2019controlled, wolff2021broadband, beutel2021detector, gyger2021reconfigurable, beutel2022cryo, haussler2023scaling, jaha2024kinetic} & \cite{beyer2015waveguide, shainline2017room, mcdonald2019iii} & \cite{haussler2020amorphous} & 73\% \cite{wolff2021broadband} \\
\hline
LNOI & \cite{sayem2020lithium} & \cite{lomonte2021single, tao2025single} & — & \cite{colangelo2024molybdenum} & — \\
\hline
LN & \cite{smirnov2018superconducting, agruzov2019superconducting} & \cite{prencipe2023wavelength} & \cite{hopker2017towards, hopker2021integrated} & — & 6\% \cite{smirnov2018superconducting} \\
\hline
GaAs & \cite{sprengers2011waveguide, sahin2013waveguide, reithmaier2015chip, digeronimo2016integration, kaniber2016integrated, schwartz2018fully} & — & — & — & 4\% \cite{sahin2013waveguide} 3.4\% \cite{sprengers2011waveguide} \\
\hline
PCD & \cite{rath2015superconducting, kahl2016high, rath2016travelling} & — & — & — & — \\
\hline
SCD & — & \cite{atikian2017novel} & — & — & — \\
\hline
Glass & \cite{hou2021waveguide} & — & — & — & 1.7\% \cite{hou2021waveguide} \\
\hline
AlN & — & \cite{najafi2015chip} & — & — & — \\
\hline
TaO & — & \cite{wolff2020superconducting} & — & — & — \\
\cline{1-6}
Max. OCDE & 99.73\% \cite{li2025surpassing} 98.9\% \cite{psiquantum2025manufacturable} & 96\% \cite{akhlaghi2015waveguide} & 90\% \cite{mcdonald2019iii} & 73\% \cite{haussler2020amorphous} & \\
\hline
\end{tabularx}
\end{table}

Table \ref{tab:optic_overview} provides a comparative overview of the key properties of major photonic materials used for integration of SNSPDs, highlighting their suitability for quantum photonic applications. In particular, the comparison considers typical propagation losses, the Kerr nonlinear coefficient determining the efficiency of photon-pair generation via nonlinear processes, the electro-optic coefficient enabling high-speed modulation, as well as integration density determined by the refractive index and the maturity of the SNSPD integration technology assessed in proportion to the number of existing studies.

\begin{table}[htbp]
\caption{Comparison of key figures of merit for selected photonic platforms}
  \label{tab:optic_overview}
  \centering
    \begin{tabularx}{\textwidth}{ 
        >{\raggedright\arraybackslash}m{2.5cm} 
        >{\centering\arraybackslash}m{1.5cm} 
        >{\centering\arraybackslash}m{1.8cm}
        >{\centering\arraybackslash}m{1.8cm} 
        >{\centering\arraybackslash}m{1.7cm} 
        >{\centering\arraybackslash}m{1.7cm}
    }
\hline
Property & SOI & SiN & LNOI & GaAs & Diamond \\
\hline
Transparency window, $\mu$m & 
\makecell{1.1--3.7 \\ \cite{badri2020coupling}} & 
\makecell{0.4--4 \\ \cite{buzaverov2024silicon}} & 
\makecell{0.35--4.5 \\ \cite{zhao2020shallow}} & 
\makecell{0.9--17 \\ \cite{skauli2003improved}} & 
\makecell{0.2--20 \\ \cite{jin2025diamond}} \\
\hline
Propagation losses, db/m & 
\makecell{6.5--200 \\ \cite{zhang2022ultralow}} & 
\makecell{0.034--30 \\ \cite{buzaverov2024silicon}} & 
\makecell{2.7--50 \\ \cite{lin2020advances}} & 
\makecell{100--200 \\ \cite{chang2018heterogeneously}} & 
\makecell{34--500 \\ \cite{hausmann2014diamond}} \\
\hline
Kerr nonlinear coefficient, $m^{2}$/W & 
\makecell{$5\times 10^{-18}$ \\ \cite{li2026pockels}} & 
\makecell{$2.5\times 10^{-19}$ \\ \cite{sun2022silicon}} & 
\makecell{$1.8\times 10^{-19}$ \\ \cite{li2026pockels}} & 
\makecell{$1.6\times 10^{-17}$ \\ \cite{baron2009light}} & 
\makecell{$1.3\times 10^{-19}$ \\ \cite{sun2022silicon}} \\
\hline
Electro-optic coefficient, pm/V & 
\makecell{$\sim0$ \\ \cite{zhu2021integrated}} & 
\makecell{$\sim0$ \\ \cite{zhu2021integrated}} & 
\makecell{30 \\ \cite{bahadori2020fundamental}} & 
\makecell{1.5 \\ \cite{zhao2010thickness}} & 
\makecell{$\sim0$ \\ \cite{zhu2021integrated}} \\
\hline
Integration density & High & Moderate & Moderate & High & Moderate \\
\hline
SNSPD integration technology & Mature & Mature & Emerging & Emerging & Emerging \\
\hline
\end{tabularx}
\end{table}

In this section, we examine implemented photonic platforms for SNSPD integration, comparing their properties, integration feasibility and demonstrated performance.

\subsubsection{Silicon-on-insulator (SOI)}

Silicon-on-insulator (SOI) photonics relies on high-index-contrast waveguides made of silicon and surrounded by silicon oxide cladding\cite{bogaerts2014silicon}. SOI-based PICs operate over a wavelength range of 1.1–3.7 $\mu$m\cite{badri2020coupling}, which makes silicon one of the preferred platforms for applications in telecommunications and optical interconnects\cite{heck2012hybrid}. For telecom wavelengths, SOI waveguides typically exhibit propagation losses of around 1 dB/cm\cite{vlasov2004losses, tran2018ultra}, while simultaneously providing strong optical confinement and compatibility with CMOS fabrication processes\cite{shekhar2024roadmapping}. Owing to these advantages, the silicon platform has become one of the most widely adopted platforms for integrating single-photon detectors.

Well-established coupling techniques for SOI waveguides have enabled SNSPDs integrated with silicon waveguides to achieve system detection efficiencies up to 20\% \cite{shu2025chip} and 40\% \cite{ono2021si} for grating and edge coupling, respectively. Buckley et al. have first demonstrated the integration of eleven SNSPDs with a cryogenic, electrically injected, waveguide-coupled Si light-emitting diodes (LED) on a single chip\cite{buckley2017all}. In a recent study by PsiQuantum \cite{psiquantum2025manufacturable}, ultrahigh performance SNSPDs have been integrated on a single chip with a broad range of key quantum photonic components, including single-photon sources based on spontaneous four-wave mixing (SFWM), low-loss silicon waveguides and thermo-optic switchers, as shown in Fig.~\ref{fig:review_materials}(b). This integration has enabled record performance, achieving Hong–Ou–Mandel interference visibility of 99.5$\pm$0.25\% and high-fidelity two-qubit operations exceeding 99\%. The study has also reported the implementation of iSNSPDs in a pseudo-photon-number-resolving (pseudo-PNR) configuration, which will be discussed in more detail in Section 5. By combining high-performance detectors with a scalable, CMOS-compatible platform, this work has demonstrated a clear pathway toward industrially viable quantum photonic processors.

While the integration of iSNSPDs with other active elements on photonic integrated circuits has been explored, SOI has also served as a testbed for alternative, non-monolithic SNSPD integration strategies. Najafi et al. has reported heterogeneous integration of ten SNSPDs fabricated on SiN membranes onto SOI waveguides, employing a micrometer-scale flip-chip process that has enabled scalable incorporation of SNSPDs into diverse photonic platforms. Detectors fabricated using this approach have achieved SDE and OCDE values of up to 19\% and 52\%, respectively \cite{najafi2015chip}. Extending idea of non-monolithic integration, Tao et al. have introduced a transfer-printing technology for integrating SNSPDs onto both silicon-on-insulator and lithium niobate on insulator (LNOI) waveguides, as shown in Fig.~\ref{fig:review_materials}(a), achieving sub-100 Hz dark count rates and OCDE around 8\% \cite{tao2025single}. Recently, Li et al. have demonstrated heterogeneous integration of transversal-design SNSPDs on a silicon photonic platform. The transversal or comb geometry with nanowire bends positioned outside the waveguide region has ensured uniform detector sensitivity along the waveguide regardless of light propagation direction. This has enabled dual-detector integration on a single waveguide, as shown in Fig.~\ref{fig:review_materials}(c), with uniform sensitivity providing self-calibration capability that has eliminated uncertainties associated with classical power measurements and fiber-to-chip coupling losses, thereby achieving 99.73\% on-chip detection efficiency at 1550 nm \cite{li2025surpassing}.
In all the cases of heterogeneous integration considered, pre-selection has allowed the transfer of only high-performance devices, thereby mitigating yield limitations. Although heterogeneous integration of single-photon detectors on PICs has not yet matured for large-scale high-yield fabrication, it represents a promising route toward incorporating state-of-the-art SNSPDs into PICs across arbitrary material platforms, including those hosting active photonic components. 

Despite its advantages, SOI photonics faces intrinsic performance limitations. First, the large index contrast between silicon and SiO$_{2}$ enhances sidewall roughness sensitivity, leading to increased propagation losses. Second, nonlinear effects in silicon restrict the maximum optical power that can be guided\cite{shekhar2024roadmapping, heck2012hybrid}.

Silicon remains one of the most mature and scalable photonic platforms, benefiting from CMOS-compatible fabrication and dense integration capabilities. Despite limited active functionality, its potential for large-scale quantum photonic circuits makes SOI one of the leading candidates for future quantum information processing systems.

\subsubsection{Silicon nitride (SiN)}

Silicon nitride (SiN) waveguide fabrication technologies have enabled a new generation of photonic integrated circuits combining ultra-low propagation losses with the ability to support both linear and nonlinear optical functionalities across a wide wavelength range from 400 to 4000 nm \cite{buzaverov2024silicon}. Over the past two decades, advances in SiN photonics have led to compact PICs with propagation losses significantly lower than those typically observed for III-V or silicon waveguides \cite{xiang2022silicon}. For telecom and visible wavelengths, losses typically reach a few dB/m \cite{chauhan2022ultra, ji2023ultra, ji2024efficient, bose2024anneal}, with record values below 0.1 dB/m \cite{puckett2021422, liu2022ultralow}. The SiN platform has also enabled resonators with quality factors on the order of tens of millions \cite{ji2016breaking, bose2024anneal, ji2024efficient}, thermo-optic (TO) modulators with bandwidths exceeding 10 kHz \cite{nejadriahi2021efficient, zeng2025submilliwatt} and grating couplers with insertion losses below 1 dB \cite{lomonte2024scalable}. However, the inherently low electro-optic coefficient of SiN (8.31$\pm$5.6 fm/V) limits its performance for active photonic elements such as high-speed modulators \cite{miller2015electro}, often necessitating hybrid integration with diverse photonic platforms to achieve functional electro-optic operation \cite{churaev2023heterogeneously}.

\begin{figure}[htbp]
\centering\includegraphics[width=13.5cm]{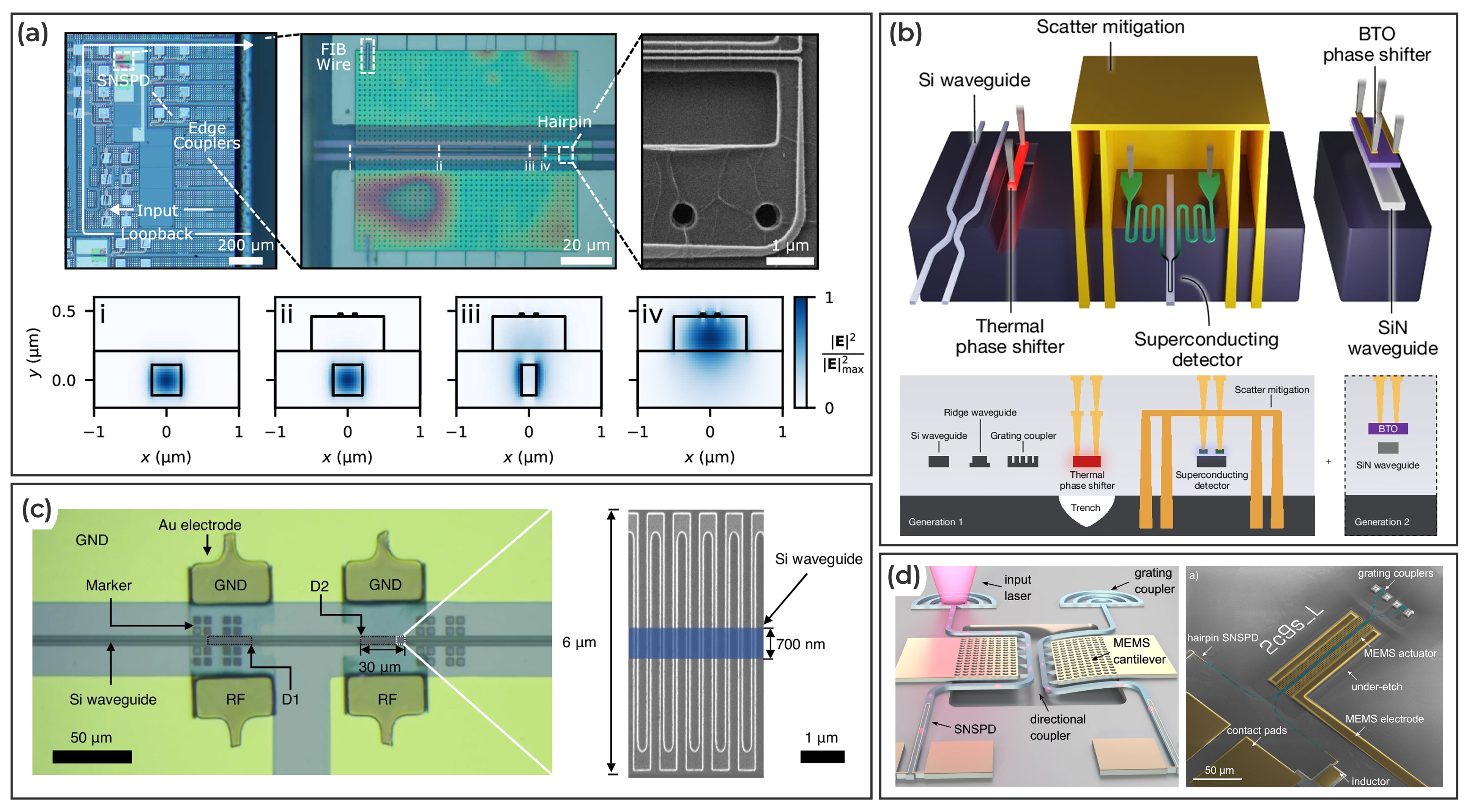}
\caption{Overview of superconducting single-photon detector on-chip integration implementations. (a) SNSPD integrated on SOI platform via transfer printing. Reprinted from \cite{tao2025single} under a Creative Commons licence. (b) Schematic representation of components integrated on a Si-based photonic platform. Adapted from \cite{psiquantum2025manufacturable} under a Creative Commons licence. (c) Two comb nanowire detectors integrated on a single waveguide, forming an in-situ self-calibrating setup on chip. Adapted from \cite{li2025surpassing} under a Creative Commons licence. (d) MEMS-based reconfigurable photonic integrated circuit with on-chip superconducting single-photon detectors. Adapted from \cite{gyger2021reconfigurable} under a Creative Commons licence.}
\label{fig:review_materials}
\end{figure}

The SiN platform has attracted the largest number of studies on waveguide-integrated SNSPDs. Single-photon detectors on silicon nitride hold the record for system detection efficiency of 73\%, as have been demonstrated by Wolff et al. \cite{wolff2021broadband} using a 3D interface based on total internal reflection (TIR) to efficiently couple large mode field diameter optical fibers to iSNSPDs. Since SiN has a negligible electro-optic coefficient, alternative approaches to on-chip reconfigurability at cryogenic temperature have been explored. One promising solution is the use of microelectromechanical (MEMS) phase shifters, which have first been co-integrated with SNSPDs on the same chip by Gyger et al., as shown in Fig.~\ref{fig:review_materials}(d) \cite{gyger2021reconfigurable}. This device has shown low-power reconfiguration at 0.1K temperature with 90 dB high-dynamic range detection of single photons and near-MHz speed. Nonetheless, this demonstration mainly has served as proof of concept and has been limited in detector performance. Beutel et al. have achieved a significant step forward by combining a MEMS-based phase shifter with a high-performance integrated detectors \cite{beutel2022cryo}. Device has exhibited a low half-wave voltage of 4.6 V at a temperature of 1.3 K, insertion loss of only 0.7 dB and OCDE of 88\%, along with a timing jitter of 12.2 ps, demonstrating the concept feasibility that offers a viable path to fully integrated photonic circuits that combine low-loss, high-speed modulators and high-performance single-photon detectors. 

Silicon nitride has also proven to be a versatile platform for on-chip integration of SNSPDs with light sources \cite{mcdonald2019iii} as well as with single-photon emitters, such as single-walled carbon nanotubes (SWCNT) \cite{khasminskaya2016fully} and quantum dot (QD) emitters \cite{gourgues2019controlled, elsinger2019integration}. 

SiN stands out as one of the most promising platforms for quantum photonics due to its ultra-low-loss waveguides, which are essential for preserving quantum states of light over extended circuits. While it lacks intrinsic active functionality, it provides an ideal low-loss backbone for interfacing with external or heterogeneously integrated single-photon sources and modulators.

\subsubsection{Lithium niobate (LN)}

Lithium niobate (LN) has long been a cornerstone material in optics due to its outstanding intrinsic properties: it offers a wide transparency window from the visible to the mid-infrared range, a large electro-optic coefficient and strong nonlinear responses. It makes LN highly attractive for modulators, frequency converters and nonlinear optical devices \cite{zhao2020shallow}. For decades, the standard technology has relied on bulk lithium niobate waveguides, fabricated by titanium diffusion or proton exchange \cite{alferness1980advances}, which typically provide propagation losses below 1 dB/cm \cite{noda1975electro, fukuma1980optical, minford1982tow} (down to 1 dB/m with optimized fabrication process \cite{li2023ultra}), although this platform suffer from weak optical confinement and require relatively large device footprints \cite{chen2022advances}. The modern LNOI (Lithium Niobate on Insulator) platform opens new opportunities: a thin film of lithium niobate bonded onto an insulator enables strong light confinement in nanophotonic waveguides \cite{hu2012lithium}. Although the propagation losses in such structures are still higher than in bulk waveguides \cite{younesi2025fabrication}, the strong mode confinement, compactness and compatibility with integrated photonic fabrication processes make LNOI a highly promising platform for high-speed modulators, microresonators and nonlinear converters. At the same time, the material has inherent drawbacks: while lithium niobate is chemically inert and challenging to etch with high fidelity, LNOI is additionally burdened by relatively high cost and brittleness, which collectively complicate large-scale deployment \cite{li2020research, kumar2024optimization}. Moreover, the restricted thermal budget of the LNOI stack imposes stringent constraints on post-fabrication processing, complicating the deposition of high-quality crystalline superconducting films and thus hindering the monolithic integration of SNSPDs. Therefore, to date, superconducting single-photon detectors have been integrated on LNOI waveguides using optimized low-temperature deposition processes for crystalline superconductors \cite{lomonte2021single, sayem2020lithium}, using amorphous materials \cite{colangelo2024molybdenum} or heterogeneous integration \cite{tao2025single}.

In the domain of lithium niobate platform, Lomonte et al. have demonstrated the first integration of SNSPDs together with an electro-optic Mach–Zehnder interferometer on the same cryogenically operated LNOI-based chip \cite{lomonte2021single}. The integrated SNSPDs has exhibited dark count rates as low as 2 Hz and on-chip detection efficiency up to 27\%. This platform has utilized low-loss waveguides (0.2 dB/cm), as well as high-speed modulation up to 1 GHz at 1.3K, offering a compelling blueprint for scalable cryo-compatible quantum photonic circuits.

Lithium niobate combines low losses with strong electro-optic effect, making it uniquely suited for fast and reconfigurable quantum circuits. Recent progress in heterogeneous integration further positions LN as a promising host for combining high-performance active and passive components within the same chip.

\subsubsection{Gallium arsenide (GaAs)}

Gallium arsenide (GaAs) is a widely used semiconductor material in conventional photonics, featuring a broad transparency window from 0.9 to 17 $\mu$m and pronounced nonlinear properties \cite{skauli2003improved}. Furthermore, GaAs is fully compatible with established III-V technologies used for fabricating lasers, quantum dots and photodetectors, making it a strong candidate for a monolithic platform in quantum photonic applications \cite{dietrich2016gaas, koester2024gaas}. Thus, gallium arsenide (GaAs) represents the platform with the largest number of studies demonstrating the integration of quantum dots on a single chip with SNSPDs, thanks to its ability to host high-quality single-photon sources with narrow spectral lines and low emission fluctuations \cite{reithmaier2015chip, kaniber2016integrated, digeronimo2016integration, schwartz2018fully}. At the same time, GaAs platform has significant limitations related to nanofabrication: etching processes for GaAs often result in sidewall roughness that enhances waveguide losses (typically on the order of a few dB/cm \cite{chang2018heterogeneously, koester2024gaas}) and limited compatibility with CMOS technology \cite{joint2020gaas, sun2019novel}. Additionally, lack of a stable native oxide for surface passivation makes GaAs surfaces more prone to defect formation and degradation compared to silicon with its robust SiO$_{2}$ passivation \cite{jacob2022surface}.

Thus, GaAs offers the rare combination of strong nonlinearities and native integration of single-photon sources such as quantum dots. This makes it particularly relevant for monolithic quantum photonic chips, where single-photon generation and detection can all be realized within a unified III-V platform.

\subsubsection{Diamond}

Diamond has emerged as an attractive platform for integrated photonics due to its wide bandgap, providing a transmission window from the ultraviolet to the far infrared wavelengths and low optical absorption \cite{jin2025diamond}. Two main material types are used: single-crystal diamond (SCD) and polycrystalline diamond (PCD). SCD offers superior crystal quality with minimal grain boundaries, leading to ultra-low optical losses of about a few dB/cm and below \cite{rahmati2021suspended, hausmann2014diamond}, while PCD, though easier and cheaper to grow over large areas, suffers from grain-boundary scattering and about an order of magnitude higher propagation losses \cite{lenzini2018diamond, liu2019design}. Additional challenges include difficulty in large-scale SCD fabrication \cite{uwihoreye2024recent}, the need for advanced etching techniques to minimize sidewall roughness \cite{apostolova2021ultrafast}. Although numerous studies have emphasized the potential for monolithic integration of SNSPDs with single-photon sources in diamond platform \cite{rath2016travelling, atikian2017novel}, this has not yet been experimentally realized. 

Therefore, diamond remains uniquely suited for photonic integration with solid-state quantum emitters (such as nitrogen-vacancy and silicon-vacancy centers), making it a promising material for future quantum networks and sensing technologies \cite{katsumi2025recent}.

\subsubsection{Other photonic platforms}

Beyond the photonic platforms discussed above, several other materials have also been explored for SNSPD integration, while others, though not yet employed for this purpose, show great promise as candidates for scalable platforms in quantum information processing. 

Borosilicate glass offers ultra-low propagation losses (below 0.1 dB/cm \cite{shi2015fiberized}) and is widely employed in low-cost photonic circuits, though its weak nonlinear and electro-optic response limits active functionalities \cite{carvalho2008borosilicate, salah2025structural}. So far, only a single study has reported the integration of SNSPDs on borosilicate glass, where femtosecond-laser-written waveguides have been employed, demonstrating the technical feasibility of this platform \cite{hou2021waveguide}.

Tantalum pentoxide (TaO) provides propagation losses of the order of several dB/cm \cite{guo2023ultra} (below 0.1 dB/cm with optimized cladding layer \cite{itoh2015low, belt2017ultra}), while allowing strong light confinement \cite{vaid2025high}, making it attractive for compact passive devices, though it lacks strong intrinsic nonlinear or electro-optic effects \cite{ryu2020improved, guo2023ultra}. So far there has been only one demonstration of SNSPD integration on a tantalum pentoxide platform \cite{wolff2020superconducting}, yet the material continues to attract significant research interest \cite{liu2025tantalum}. 

Aluminum nitride (AlN) combines a wide bandgap with intrinsic second order nonlinearity \cite{terrasanta2022aluminum}, piezoelectric properties \cite{haider2023review} and CMOS compatibility \cite{sandeep2025impact}, enabling electro-optic modulators and nonlinear devices \cite{guo2023ultra}, while ensuring propagation losses of less than 1~dB/cm \cite{singh2024sputtered, sundarapandian2025optical}. A hybrid integration approach has been previously demonstrated by combining an AlN photon-pair source based on spontaneous parametric down-conversion (SPDC) with SiN-integrated iSNSPDs, showcasing the potential of AlN for scalable quantum photonic circuits \cite{guo2017parametric}.

Barium titanate (BTO) is a highly promising electro-optic platform for integrated quantum photonics, offering an exceptionally high Pockels coefficient (over 1000 pm/V), more than an order of magnitude above that of lithium niobate \cite{gaur2023comparative}. The implemented BTO electro-optic modulators heterogeneously integrated via oxide bond with SiN-based photonic platform, which can combine high-quality single-photon sources and iSNSPDs, have demonstrated an overall low device insertion loss of 0.1 dB, a loss-voltage product of 0.33$\pm$0.02 dB$\cdot$V and a bandwidth of over 6 GHz \cite{psiquantum2025manufacturable}. These metrics underscore BTO potential to serve as a foundational material for large-scale, fast and low-loss reconfigurable photonic networks necessary for fault-tolerant photonic quantum computing.

\subsection{Summary}

The considered photonic platforms differ significantly both in their optical properties and in the ability to provide active functionality. In Table \ref{tab:integration} we present studies demonstrating the integration of active photonic components, such as classical light sources, single-photon emitters and reconfigurable photonic elements, together with iSNSPDs on diverse photonic platforms. This overview emphasizes not only the technical feasibility of combining detectors with functional photonic circuits but also highlights the ongoing research activity and the promising potential of different material platforms for multifunctional photonic systems.

\begin{table}[htbp]
    \caption{State-of-the-Art in monolithic (M), heterogeneous (HG) and hybrid (HB) integration of iSNSPDs based on diverse superconductors with active photonic components on various platforms}
    \label{tab:integration}
    \centering
    \begin{tabularx}{\textwidth}{ 
        >{\raggedright\arraybackslash}p{3.7cm} 
        >{\centering\arraybackslash}p{1.2cm} 
        >{\centering\arraybackslash}p{1.3cm} 
        >{\centering\arraybackslash}p{2cm} 
        >{\centering\arraybackslash}p{1.7cm} 
        >{\centering\arraybackslash}p{1.1cm}
    }
        \hline
        Ref. & Platform & Light source & Single-photon source & Modulator & SNSPD \\
        \hline
        Reithmaier et al.\cite{reithmaier2015chip} & GaAs & — & QD (M) & — & NbN \\
        \hline
        Digeronimo et al.\cite{digeronimo2016integration} & GaAs & — & QD (M) & — & NbN \\
        \hline
        Kaniber et al.\cite{kaniber2016integrated} & GaAs & — & QD (M) & — & NbN \\
        \hline
        Khasminskaya et al.\cite{khasminskaya2016fully} & SiN & — & SWCNT (HG) & — & NbN \\
        \hline
        Guo et al.\cite{guo2017parametric} & AlN/SiN & — & SPDC (HB) & — & NbTiN \\
        \hline
        Buckley et al.\cite{buckley2017all} & SOI & LED (M) & — & — & WSi \\
        \hline
        Schwartz et al.\cite{schwartz2018fully} & GaAs & — & QD (M) & — & NbN \\
        \hline
        Gourgues et al.\cite{gourgues2019controlled} & SiN & — & QD (HG) & — & NbTiN \\
        \hline
        McDonald et al.\cite{mcdonald2019iii} & SiN & LED (M) & — & — & WSi \\
        \hline
        Elsinger et al.\cite{elsinger2019integration} & SiN & — & QD (HG) & — & NbN \\
        \hline
        Lomonte et al.\cite{lomonte2021single} & LNOI & — & — & EO (M) & NbTiN \\
        \hline
        Gyger et al.\cite{gyger2021reconfigurable} & SiN & — & — & MEMS (M) & NbTiN \\
        \hline
        Beutel et al.\cite{beutel2022cryo} & SiN & — & — & MEMS (M) & NbTiN \\
        \hline
        PsiQuantum\cite{psiquantum2025manufacturable} & SOI & — & SFWM (M) & TO (M) & NbN \\
        \hline
    \end{tabularx}
\end{table}

\begin{figure}[htbp]
\centering\includegraphics[width=13cm]{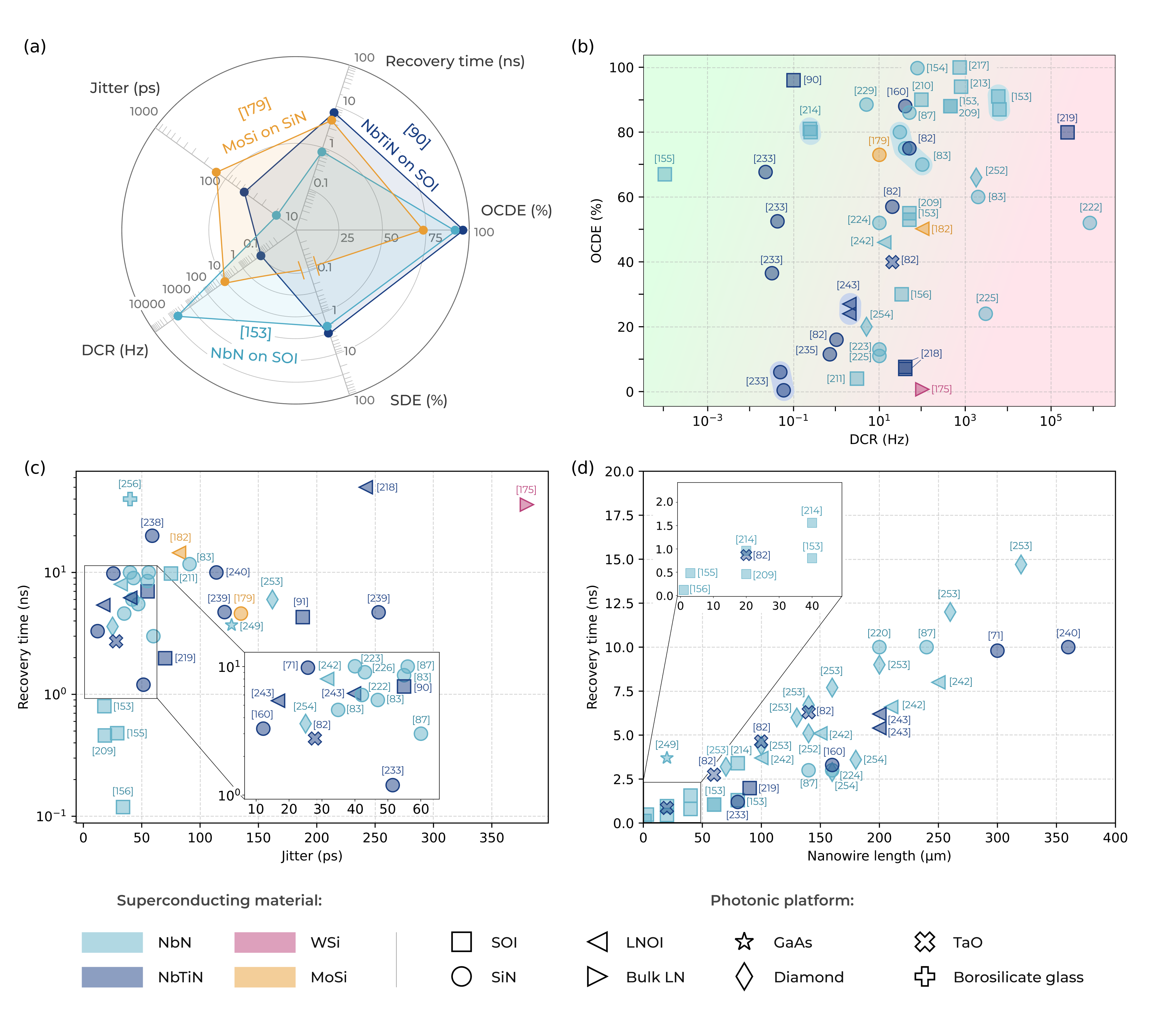}
\caption{Overview of achieved performance metrics of iSNSPDs based on various superconducting materials and on different photonic platforms. (a) Performance metrics of state-of-the-art iSNSPDs, achieved within a single device. (b) Performance landscape plotting on-chip detection efficiency at telecom wavelengths against dark count rate for state-of-the-art iSNSPDs. The gradient from red to green corresponds to the trend of the characteristic improving from poor to favorable values. (c) Recovery time and timing jitter performance space for state-of-the-art iSNSPDs. (d) Correlation between recovery time and nanowire length for iSNSPDs based on crystalline superconductors.}
\label{fig:plots}
\end{figure}

To provide a comprehensive overview of the state-of-the-art in integrated SNSPDs, we performed a systematic analysis of performance metrics reported across the studies. While record performance metrics have been discussed above for detectors based on various materials, the simultaneous achievement of high detection efficiency, low dark count rate and advanced timing performance remains challenging. Fig.~\ref{fig:plots}(a) summarizes the performance metrics of state-of-the-art detectors based on different superconducting materials, achieved within a single device. Based on a comprehensive analysis of published works on iSNSPDs, Fig.~\ref{fig:plots}(b-d) shows three representative correlations, each highlighting a specific aspect of detector operation and scalability: detection efficiency, noise level, temporal response and the impact of device geometry. 

In Fig.~\ref{fig:plots}(b), the dependence of on-chip detection efficiency at telecom wavelengths on dark count rate is shown. This relation is crucial as it captures the balance between achieving high sensitivity and suppressing noise. Reported data generally illustrate that increasing OCDE often leads to higher DCR in the range of about 10 to 1000 Hz, although several state-of-the-art devices mostly based on crystalline superconductors demonstrate the ability to combine efficiencies above 60\% with sub-Hz DCR. 

In Fig.~\ref{fig:plots}(c), recovery time is plotted against timing jitter, reflecting the temporal dynamics of the detectors. This comparison is important since it links the count rate of detector with its temporal resolution. This comparison clearly shows that iSNSPDs based on amorphous superconducting materials have not yet achieved the same high temporal performance as those based on crystalline ones. It may be related both to the fundamental properties of the materials and to the number of studies with detectors based on them. Most devices cluster within the range of sub-10 ns recovery time and below 60 ps jitter, while the best-performing demonstrations achieve jitter below 20 ps at sub-ns recovery times. A noticeable correlation can be seen in the recovery time versus jitter dependence: devices with longer recovery generally exhibit higher timing jitter. This behavior can be attributed to the fact that most of the relevant factors, such as nanowire geometry, material properties and operation conditions, simultaneously affect both characteristics in a comparable manner, leading to their concurrent increase or reduction, as discussed in Section 3. 

Fig.~\ref{fig:plots}(d) compiles reported recovery times versus nanowire length from various experimental studies. It is worth noting that the figure does not show detectors with additional inductance connected in series. For both NbN and NbTiN devices, the dependence is close to linear and the two datasets exhibit almost indistinguishable slopes, although NbTiN generally shows slightly better recovery due to lower kinetic inductance. The observed deviations from linearity can largely be attributed to variations in nanowire width, thickness and operating conditions reported in different studies, all of which influence the detection process dynamics.

Further progress in iSNSPD-based devices performance will largely depend on the advancement of photonic platforms, particularly in terms of propagation losses minimization while enabling complex on-chip architectures. A major focus is on refining fabrication techniques and developing heterogeneous integration strategies to combine distinct material systems in a single device. Such strategies hold the potential to merge low-loss waveguides with materials offering enhanced active functionalities, thereby opening new routes toward scalable quantum information processing systems.

\section{Photon-number resolution}

Time- and number-resolved photon detection is critical for quantum information processing\cite{lita2022development}. While SNSPDs have predominantly been employed as binary detectors (distinguishing only between zero and $\geq $1 photons), recent advances now enable photon-number resolution (PNR) by exploiting intrinsic detector response characteristics. 

Intrinsic PNR methods utilize the physical dependence of SNSPD output signals (pulse amplitude or rise time) on incident photon number\cite{schapeler2025optimizing}. An alternative pseudo-PNR strategy involves using multiple independent detector sections. Key PNR approaches for superconducting single-photon detectors are summarized in Table \ref{tab:pnr_methods}, highlighting existing implementations for integrated detectors and the maximum number of photons $N_{max}$ that has been resolved by iSNSPDs using various methods.

\begin{table}[htbp]
    \caption{Comparison of PNR methods and implementations}
    \label{tab:pnr_methods}
    \centering
    \begin{tabularx}{\textwidth}{ 
        >{\raggedright\arraybackslash}m{1cm} 
        >{\raggedright\arraybackslash}m{1.5cm} 
        >{\raggedright\arraybackslash}m{1.5cm} 
        >{\centering\arraybackslash}m{1.2cm}
        >{\centering\arraybackslash}m{1.4cm} 
        >{\centering\arraybackslash}m{1.8cm} 
        >{\centering\arraybackslash}m{2cm}
    }
        \hline
        Type & PNR method & Measured parameter & SDE depends on $N$ & $N_{max}$ & Integrated realizations & Stand-alone realizations \\
        \hline
        
        \multirow{3}{*}{\parbox[c]{1cm}{Pseudo-PNR}} &  \multirow{2}{*}{\parbox[c]{1.5cm}{\raggedright Nanowire segmentation}} & Pulse amplitude & Yes & 20 \cite{page2025scalable} & \cite{gaggero2019amplitude, sahin2013waveguide, psiquantum2025manufacturable, page2025scalable} & \cite{divochiy2008superconducting, cheng2012photon, schmidt2019characterization, mattioli2015photon, dryazgov2023micrometer, stasi2023fast, stasi2024enhanced, hao2024compact, liu2014photon, tao2019high, mattioli2016photon} \\
        \cline{3-7}
        & & Pulse delay time & Weakly & 100 \cite{cheng2023100} & \cite{cheng2023100} & \cite{oripov2023superconducting, schapeler2022information} \\
        \cline{2-7}
        & \parbox[c]{1.5cm}{\raggedright Inline detection} & Number of events & Yes & 2 \cite{martinelli2026inline} & \cite{martinelli2026inline} & — \\
        \hline
        
        \multirow{6}{*}{\parbox[c]{1cm}{Intrinsic PNR}} 
& \multirow{2}{=}{\parbox[c]{1.5cm}{\raggedright Impedance matching}} 
    & Pulse amplitude & No & — & — & \cite{zhu2019superconducting, zhu2020resolving, colangelo2023impedance, kong2025superconducting} \\
\cline{3-7}
& & Rising edge time & No & — & — & \cite{davis2022improved} \\
\cline{2-7}
& \multirow{2}{=}{\parbox[c]{1.5cm}{\raggedright Pulse noise reduction}} 
    & Pulse waveform & No & — & — & \cite{endo2021quantum, sempere2022reducing} \\
\cline{3-7}
        & & Pulse amplitude & No & — & — & \cite{cahall2017multi, hao2021improved} \\
        \cline{2-7}
        & Jitter reduction & Rising edge time & No & 3 \cite{jaha2024kinetic} & \cite{jaha2024kinetic} & \cite{sauer2023resolving, schapeler2024electrical, los2024high, sidorova2025jitter} \\
        \cline{2-7}
        & Rise time increase & Rising edge time & No & 3 \cite{jaha2024kinetic} & \cite{jaha2024kinetic} & \cite{kong2024large} \\
        \hline
    \end{tabularx}
\end{table}

Below, each of the listed methods will be considered in more detail with an emphasis on the operating principle, integration possibilities with iSNSPDs and existing trade-offs.

\subsection{Pseudo-photon-number-resolution}

The first demonstration of photon-number resolution using a superconducting nanowire single-photon detector has been achieved by dividing the nanowire into multiple sections\cite{divochiy2008superconducting}. In pseudo-PNR (or quasi-PNR) schemes, the sensitive element of the detector is typically segmented into several serially connected sections, each section is shunted by an on-chip resistor with a resistance lower than that of the nanowire in the normal state. When a photon is absorbed by one of the sections, its state switches from superconducting to resistive and the bias current is diverted into the corresponding resistor, while simultaneous absorption in multiple sections redistributes the current across several resistors, making the detector output pulse scale with the number of absorbed photon. This method remains the most widely used approach for demonstrating PNR in both standalone and integrated SNSPDs. In the latter case, detector sections are typically patterned on the same waveguide. However, alternative implementations place SNSPD sections on different waveguides, using a Y-splitter to distribute photons between two paths\cite{gaggero2019amplitude,page2025scalable}, or arrange multiple independent detectors inline on a single waveguide\cite{martinelli2026inline}. Waveguide-integrated SNSPDs segmented with parallel resistors have demonstrated resolution of up to 16 photons\cite{schutte2023waveguide}. Nevertheless, despite the simplicity of implementation and readout, this approach faces inherent limitations. When more than one photon is absorbed within a single section, the information about photon number is lost, as in the case of a conventional SNSPD. As a result, the detection efficiency of such a system is determined by the photon number $N$. Consequently, if an application requires photon-number-independent efficiency, the detector must be segmented into a number of sections significantly larger than the maximum photon number to be resolved in the experiment. The scalability of this method is fundamentally constrained by three factors: the capability of the readout electronics to resolve closely spaced voltage levels, the physical size of the resistive elements and the thermal budget, which must permit high-performance operation at cryogenic temperatures.

An alternative architecture for pseudo–PNR has been demonstrated for a waveguide-integrated SNSPD by Cheng et al.\cite{cheng2023100}. In this approach, which has also been based on segmenting the nanowire, photon-number resolution has been achieved through spatiotemporal multiplexing. The nanowire in this design has been divided into 100 serially connected superconducting sections, separated by delay lines and reset loops. Since the response of an SNSPD generates two voltage pulses of opposite polarity, both the number of detected photons, determined from the total number of voltage peaks on the time axis, and the number of the triggered sections, identified by the temporal position of those peaks, have been inferred. Importantly, unlike previous pseudo-PNR implementations based on amplitude multiplexing, the detector sections here have been made sufficiently small to make the probability of multiple-photon absorption within a single section negligible, thereby minimizing the associated counting error and strongly suppressing dependence of the system detection efficiency on the number of photons.

\subsection{Intrinsic photon-number resolution}

Fundamental physical principles of SNSPD operation suggest that the detector response should inherently vary with the number of resistive domains formed in the superconducting nanowire, thereby providing intrinsic photon-number resolution. Nevertheless, this variation remains unobserved in conventional devices due to several limiting factors. Therefore, significant efforts are presently directed toward realizing the intrinsic PNR capability rooted in the fundamental detector physics.

\subsubsection{Impedance matching}

A promising route toward intrinsic photon-number resolution in SNSPDs relies on integrating on-chip impedance-matching structures\cite{zhu2019superconducting, zhu2020resolving, colangelo2023impedance, kong2025superconducting}, designed to bridge the intrinsic impedance of the nanowire, typically a few hundred to several thousand ohms, to the 50$\Omega$ impedance of standard readout electronics. By employing impedance tapers, the SNSPD output amplitude becomes sensitive to the number of photon-induced hotspots, enabling photon-number resolution via amplitude multiplexing.

Although this method can be extended to integrated detectors, photon-number resolution based solely on impedance matching has not yet been demonstrated in iSNSPDs. The approach is appealing due to its purely planar geometry and the lack of additional readout components. However, the output amplitude grows sublinearly with photon number, which limits the resolvable photon range and leads to a high probability of state misclassification, manifested as low fidelity\cite{zhu2019superconducting, colangelo2023impedance}. Furthermore, achieving effective impedance matching requires taper lengths on the millimeter scale, even with optimized design and layer stacks, representing a severe bottleneck for system scalability\cite{zhu2020resolving}.

\subsubsection{Pulse noise reduction}

In addition to the impedance mismatch between the nanowire and the readout electronics, the indistinguishability of output pulse amplitudes at different photon numbers is worsened by electrical noise, which primarily originates in the room-temperature amplifier chain. Thereafter, an alternative strategy for implementing PNR is to add a cryogenic amplifier directly coupled to the detector output, thereby minimizing added noise and allowing the pulse amplitudes corresponding to different numbers of photons to be distinguished\cite{cahall2017multi, cahall2018scalable, sempere2022reducing}. Alternatively, Endo et al.\cite{endo2021quantum} have optimized such a cryogenic amplifier for broadband signal profiles, which enabled photon-number resolution through waveform analysis rather than amplitude readout. 

Photon number resolution by reducing iSNSPD pulse amplitude noise has not been demonstrated, although the use of cryogenic amplifiers has subsequently allowed the development of a new approach for intrinsic PNR for both stand-alone and integrated detectors, which will be discussed below. While this approach does not require changes in the design of the detector itself, it comes with significant trade-offs. Each PNR detector in the system requires its own dedicated cryogenic amplifier, dramatically increasing the cost of the experimental setup. Scalability is further constrained by the practical limitations of cryogenic platforms: all amplifiers must reside at the same cryostat stage, where physical space, thermal budget and the number of available electrical feedthroughs are inherently limited.

\subsubsection{Jitter reduction and rise time increase}

While exploring photon-number resolution by reducing pulse noise, it has been observed that not only the output-pulse amplitude but also the pulse rise time exhibits a dependence on the number of photons\cite{endo2021quantum}. Because the rise time of the SNSPD output signal decreases with photon number, the same voltage amplitude corresponding to the trigger level of the readout electronics is reached at different delays relative to photon arrival. In practice, however, the relatively large timing jitter of SNSPDs, both in most experimental demonstrations and in commercial systems, has masked this effect, rendering the $N$-dependence negligible compared to the intrinsic temporal instability of the device. Later studies revealed that once detector jitter is reduced, the photon-detection histogram develops a comb-like structure, with distinct peaks corresponding to different photon numbers.

Building on this idea, rise-time differences have been exploited to demonstrate intrinsic photon-number resolution in standalone SNSPDs \cite{los2024high, kong2024large}, including commercially available devices after modification \cite{sauer2023resolving, schapeler2024electrical, sidorova2025jitter}. Observable rise-time dependence on photon number has been achieved through two routes: (1) reducing detector jitter to produce sufficiently narrow temporal peaks that can be clearly distinguished\cite{sauer2023resolving, schapeler2024electrical, los2024high, sidorova2025jitter} and (2) intentionally increasing the kinetic inductance to extend the rise time and separate the peaks in the temporal domain\cite{kong2024large}. Unlike waveform-based methods that require post-processing, these approaches allow real-time photon counting, either by measuring the delay between photon arrival (by readout of the synchronized pulse from the photon source) and rising edge of the detector voltage pulse \cite{sauer2023resolving}, or by indirectly estimating the rise time through splitting the output signal and feeding it to two independent time-tagging channels with different trigger thresholds\cite{kong2024large}. Recently, Jaha et al. have reported the first demonstration of intrinsic PNR in integrated SNSPD\cite{jaha2024kinetic}. Using a cryogenic amplifier, they have systematically investigated the impact of enhanced nanowire kinetic inductance and reduced system jitter on the PNR fidelity. 

This approach offers two key advantages: high fidelity and strong scalability potential, particularly for the inductance-based approach, which can be realized by integrating compact kinetic-inductance elements either on-chip or in the readout circuit\cite{page2025scalable}. Nonetheless, practical limitations remain. Achieving high performance requires either cryogenic amplifiers, which substantially increase system cost and hinder scalability, or additional kinetic inductance, unavoidably limiting detector count rate.

\subsection{Summary}

A key limitation of most existing PNR architectures is the trade-off between fidelity, scalability and system complexity. Multi-section SNSPDs with resistive elements suffer from a fundamental loss of photon-number information whenever multiple photons are absorbed within the same section, constraining efficiency at higher photon numbers\cite{schutte2023waveguide}. Approaches based on impedance matching and amplitude multiplexing enable straightforward on-chip implementations but exhibit sub-linear amplitude scaling, ultimately limiting the resolvable photon number\cite{zhu2019superconducting,colangelo2023impedance}. Waveform- and rise-time-based schemes offer high fidelity and scalability, yet they critically depend on ultra-low-jitter performance and often require cryogenic amplifiers. From an engineering standpoint, achieving high-yield fabrication of uniform nanowire arrays, reproducible impedance tapers, or long kinetic-inductance elements remains a non-trivial challenge, particularly in complex photonic circuits. Finally, thermal budgets and the limited number of cryostat feedthroughs impose additional constraints on large-scale deployment.

In summary, while PNR-iSNSPDs are still at an early stage of development, their rapid evolution demonstrates strong potential. Overcoming the challenges of fidelity, scalability and cost will likely define the next decade of research, ultimately positioning PNR-SNSPDs as indispensable components in the global roadmap toward large-scale quantum photonic technologies.

\section{Challenges and Outlook}

Although substantial progress has been achieved in waveguide-integrated superconducting single-photon detectors, a number of challenges related to materials, device architectures and integration approaches continue to limit performance and scalability of WSNSPS-based systems.

Historically, "waveguide-top" traveling-wave SNSPDs, where the superconducting wire is patterned directly atop the waveguide, have established the baseline for high performance in integrated photonics. This architecture remains technologically straightforward due to its minimal layer stack and direct patterning approach, making it the most prevalent implementation to date. However, it suffers from a fundamental trade-off between absorption efficiency, recovery time and waveguide propagation losses.

A central architectural tension in integrated SNSPDs is the need to maximize modal overlap with the nanowire for high absorption efficiency while preserving the ultra-low propagation losses demanded by large-scale photonic circuits. Conventional architecture ensures strong interaction between the optical mode and the nanowire, enabling near-unity absorption efficiency, but the absence of a cladding layer increases waveguide propagation losses, complicating co-integration with ultra-low-loss routing optics \cite{li2021design}. To address this challenge, two distinct architectural strategies have emerged. The first involves forming the cladding layer on top of the nanowire, which is directly interfaced with the guided mode. Although a comprehensive demonstration of both low propagation losses and high detector performance has not yet been performed, this design has been experimentally demonstrated in several studies, confirming both its feasibility and its potential for reducing excess waveguide loss while preserving strong light-nanowire interaction \cite{hu2011efficient, rhazi2022development,buckley2020integrated,psiquantum2025manufacturable}. This approach allows reducing the propagation losses without degrading the absorption efficiency, however, it introduces two fabrication challenges. First, additional etching process is required to selectively remove the cladding material and expose the detector contact pads for electrical interconnection, complicating the fabrication technology \cite{buckley2020integrated}. Second, the elevated temperature deposition of common cladding materials (e.g., silicon dioxide or silicon oxynitride) can degrade the superconducting properties of the underlying nanowire \cite{hatano1988effects, yang2005fabrication}. The second approach, which involves fabricating the SNSPD on top of the cladding layer, remains theoretical to date, with existing study providing simulation-based analyses of its feasibility \cite{li2021design}. Placing the nanowire on a deposited upper cladding protects waveguide loss budgets without the need for additional operations after the detector fabrication, allowing the use of the substantive PIC fabrication technology. However, this architecture weakens field overlap, so high absorption then requires longer nanowires, which increases kinetic inductance and slows response dynamics. Thus, careful calculation of the cladding layer thickness and the length of the nanowire required for high performance needs to be done \cite{li2021design}.

The fundamental nature of this absorption–speed–loss triad represents one of the key tradeoffs of integrated SNSPDs: shortening the nanowire reduces kinetic inductance and improves temporal response, but also reduces absorption unless the optical mode is engineered to compensate \cite{hu2009efficiently}. To mitigate these trade-offs, several other architectural strategies have emerged.

An important class of alternative integrated SNSPD architectures departs from the travelling-wave absorber paradigm by embedding the superconducting nanowire in a resonant photonic structure. Resonant integration, implemented as two- (2D) \cite{yun2020superconducting} or one-dimensional (1D) \cite{akhlaghi2015waveguide, vetter2016cavity} photonic crystal cavities and microring or racetrack resonators \cite{tyler2016modelling, villarreal2016modelling}, permits strong local field enhancement and effective optical interaction lengths, ensuring near-unity absorption efficiency with recovery times well below those of conventional travelling-wave devices. A fundamental limitation of resonant enhancement architectures is their intrinsic narrowband spectral response, however, this very property of extreme wavelength selectivity may be advantageous for applications targeting specific quantum emitters or fixed-wavelength quantum protocols, presenting a powerful strategy for achieving high-performance single-photon detection at predetermined wavelengths.

\begin{figure}[htbp]
\centering\includegraphics[width=13.5cm]{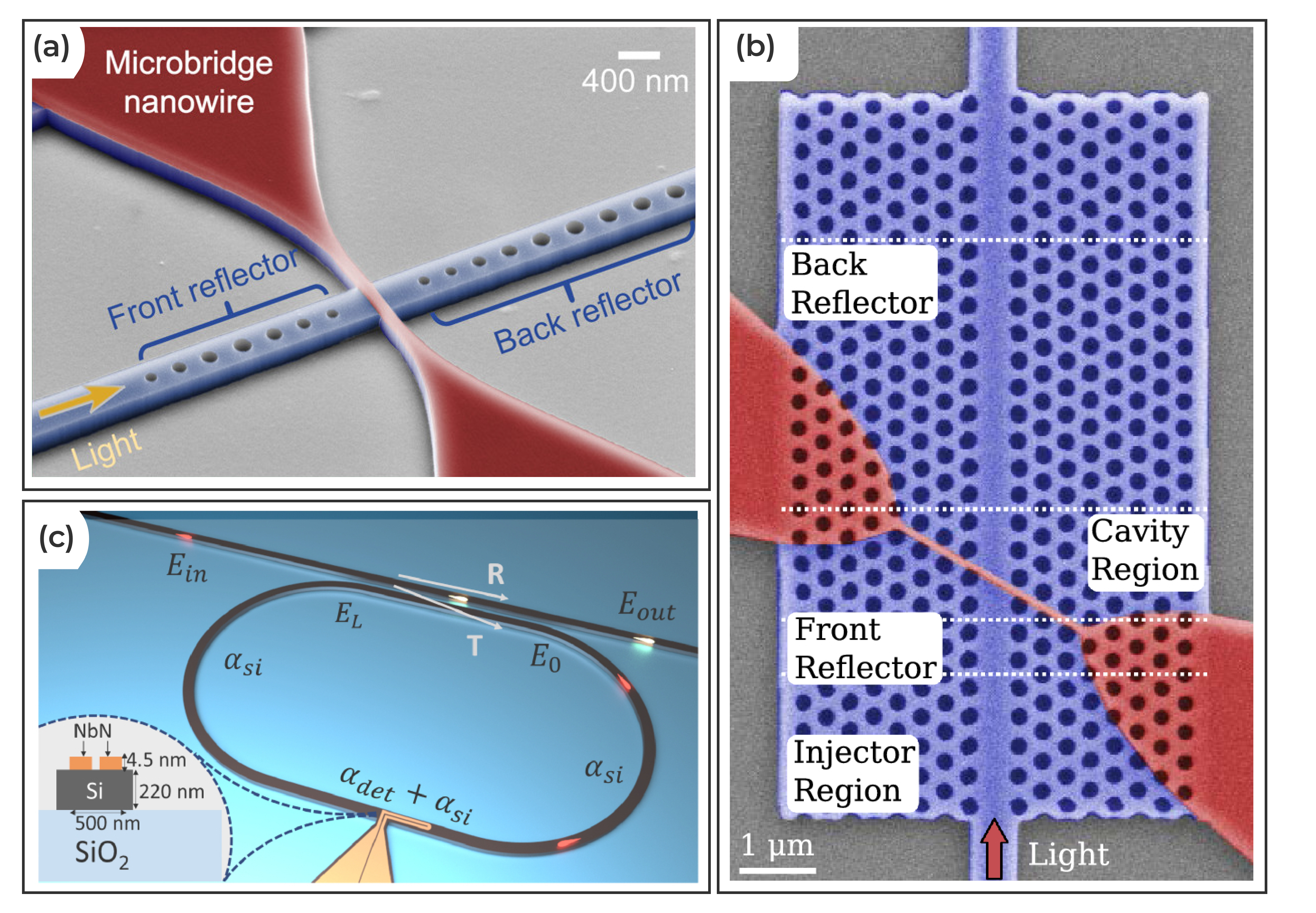}
\caption{Overview of resonant structure integration into the iSNSPD-based systems (a) Microbridge-shape iSNSPD embedded in 1D photonic crystal cavity. Reprinted with permission from \cite{vetter2016cavity} \copyright 2016 American Chemical Society. (b) iSNSPD embedded in 2D photonic crystal cavity. Reprinted with permission from \cite{munzberg2018superconducting} \copyright 2018 Optical Society of America. (c) Schematic of iSNSPD integrated in racetrack resonator. Reprinted from \cite{tyler2016modelling} under a Creative Commons licence}
\label{fig:review_resonant}
\end{figure}

Photonic crystals (PhCs) provide periodic dielectric structures with engineered photonic bandgaps and localized defect states that strongly confine electromagnetic fields \cite{englund2005general}. When a superconducting nanowire is positioned in the high-field region of a PhC cavity, the optical absorption per unit length can be enhanced by orders of magnitude compared with a travelling-wave geometry. The approach considered has been called the coherent perfect absorber (CPA) architecture \cite{chong2010coherent}. The first integration of iSNSPD into a photonic crystal structure has been demonstrated by Akhlaghi et al.\cite{akhlaghi2015waveguide}. This pioneering architecture has implemented a 1D photonic crystal cavity formed by two photonic bandgap mirrors — precisely etched series of holes on either side of the nanowire. Thus, the absorption of incident light entering this microcavity, with arbitrarily short nanowire, has been limited by the reflectivity of the mirrors. This study has achieved a landmark near-unity OCDE, demonstrating a dramatic improvement over the near 20\% OCDE of cavity-less detectors on the same platform. This performance enhancement, however, has allowed to achieve a recovery time of only 7 ns, resulting from the intentional use of an external inductor in the readout circuit for ensuring stable, low-noise operation by facilitating an overdamped bias current recovery, thereby suppressing afterpulses and amplifier oscillations. Further study by Vetter et al. has advanced this architecture by integrating a microbridge nanowire directly across the waveguide, embedded within a 1D photonic crystal cavity, optimizing the nanowire length to minimize kinetic inductance, as shown in Fig.~\ref{fig:review_resonant}(a) \cite{vetter2016cavity}. While the resulting on-chip detection efficiency has been moderate (30\%) because of the subwavelength dimensions of the nanowire, the strategy has enabled a record-breaking 1/$e$ recovery time of under 120 ps, highlighting the possibility of overcoming the fundamental trade-off between absorption efficiency and detector count rate using resonant structures. To overcome the inherent limitation of absorption efficiency while preserving ultrafast temporal response, M{\"u}nzberg et al. have pioneered the integration of a microbridge nanowire within a two-dimensional photonic crystal cavity, as shown in Fig.~\ref{fig:review_resonant}(b) \cite{munzberg2018superconducting}. This advanced architecture has simultaneously ensured a markedly improved OCDE of up to 67\% while maintaining a sub-500 ps recovery time. This result has successfully demonstrated a more favorable compromise between efficiency and response dynamics, while optimal designs of photonic crystals for both standalone and integrated SNSPDs continue to be explored \cite{yun2020superconducting, xiao2022superconducting, xiao2023ultralow}.

Microring and racetrack resonators offer another pathway to absorption enhancement in iSNSPDs. Unlike PhC, which rely on bandgap engineering and localized defect modes, ring resonators utilize continuous total internal reflection to circulate light within a closed waveguide loop. Following the approach proposed by Tyler et al., placing a short superconducting nanowire section on top of the resonator, as shown in Fig.~\ref{fig:review_resonant}(c), could enable repeated interaction between the circulating power and the absorber, providing near-unity absorption at resonance despite nanowire lengths of only a few micrometres \cite{tyler2016modelling}. While numerous studies have successfully demonstrated co-integrated ring resonators and iSNSPDs on a single chip \cite{schuck2013matrix, shainline2017room}, the more advanced approach of directly patterning the superconducting nanowire as an integral part of the resonator remains theoretical to date \cite{tyler2016modelling, villarreal2016modelling, rhazi2022development}. 

Looking forward, iSNSPD architectures that combine resonant perfect-absorption concepts with fabrication flow that protect passive waveguide loss appear most promising for scaling. The key open problem remains system-level co-optimization: sustaining low propagation loss in complex circuits with dense detector placement, while simultaneously achieving high absorption efficiency and sub-ns recovery, providing GHz detection rates. Addressing this will likely rely on cavity enhancement to keep nanowires short, careful stack and cladding engineering to suppress excess absorption and scatter, as well as on heterogeneous integration techniques to effectively combine the advantages of various photonic platforms and high-performance detectors.

Scaling iSNSPDs to multi-channel arrays essential for emerging quantum photonic systems introduces new hurdles. Device-to-device variability in critical current, efficiency and temporal performance, often stemming from nanometer-scale film thickness or stoichiometry fluctuations, degrades array yield and uniformity \cite{yang2025superconducting}. Moreover, achieving reproducible performance across large wafers requires both process control at the fabrication level and sophisticated material characterization \cite{gao2025pixels}. Although some superconducting materials, including NbN, NbTiN, WSi and MoSi, have already been widely applied for iSNSPDs, the lack of stable, high-yield fabrication of uniform nanowires on diverse platforms impedes the transition from laboratory prototypes to scalable photonic chips\cite{steinhauer2021progress}.

Beyond device physics, integrating iSNSPD arrays into practical systems poses engineering constraints. The number of optical and electrical input and output lines in traditional cryostats becomes impractical beyond a few dozen channels, which complicates system design and significantly increases the thermal load \cite{krinner2019engineering,dang2024advances}. Efficient multiplexing schemes, such as time-division \cite{zhu2018scalable} or frequency-domain multiplexing \cite{doerner2017frequency, sinclair2019demonstration, sypkens2024frequency} and cryo-compatible readout electronics \cite{miyajima2018high, cahall2018scalable} are vital to limit cabling, control wiring complexity and minimize thermal load, but their practical implementation in SNSPD-based systems remains challenging\cite{gao2025pixels}.

Looking forward, further advances are expected to arise from three main directions. First, advancements in both materials science, including amorphous superconductors and new material platforms \cite{yadav2024evaluation,jiang2024superconducting,metuh2025toward,zugliani2025single}, and post-processing methods \cite{strohauer2023site} are likely to enable both higher performance and yield. Second, progress in heterogeneous integration and hybrid photonic platforms will be crucial for combining low-loss passive components and active functionalities such as high-speed modulation \cite{mao2022heterogeneous,rahman2025high} and entangled photon generation \cite{chen2024ultralow}, along with high-performance single-photon detection. Third, system-level innovations, including cryogenic electronics \cite{niu2022wideband}, scalable multiplexed readout \cite{gao2025pixels} and photonic–electronic co-design \cite{kramnik2025scalable} will not only be necessary for large-scale deployment of multi-detector arrays, but will also open the way to the intrinsic ability of SNSPDs to resolve the number of photons without degrading performance metrics\cite{schapeler2025optimizing}.

\section{Conclusion}

Integrated superconducting nanowire single-photon detectors represent a transformative technology for integrated quantum photonics, offering a unique combination of characteristics. Recent advancements have demonstrated exceptional performance benchmarks, including on-chip detection efficiency exceeding 99\% at telecom wavelengths \cite{psiquantum2025manufacturable,li2025surpassing}, dark count rates as low as 0.0001 Hz \cite{munzberg2018superconducting}, sub-500-ps recovery time \cite{vetter2016cavity, schuck2013matrix, munzberg2018superconducting}, timing jitter below 15 ps \cite{beutel2022cryo, schutte2023waveguide}, the ability to resolve up to 100 photons \cite{cheng2023100} and intrinsic photon-number-resolving capabilities \cite{jaha2024kinetic}. The progress in coupling techniques has made it possible to realize iSNSPDs achieving system detection efficiencies as high as 73\% together with a broad spectral range \cite{wolff2021broadband} and to demonstrate arrays containing up to 64 detectors integrated on a single chip \cite{haussler2023scaling}. Moreover, recent works have reported the integration of SNSPDs on a single photonic platforms with a range of acvive components, including light sources, single-photon emitters and high-speed modulators. These achievements underscore the potential of iSNSPD to become the cornerstone of future quantum information processing systems.

Taken together, these results highlight not only the remarkable progress in individual performance metrics, but also the increasing system-level maturity of iSNSPDs, where detector performance is now closely coupled with photonic integration and functional complexity. This evolution motivates a more structured assessment of the underlying materials, device architectures and integration strategies.

In summary, this review has systematically analyzed the current state of integrated superconducting nanowire single-photon detectors, covering their operational principles, performance metrics, material platforms, integration methodologies, challenges and opportunities. This review is intended to serve as a valuable reference for researchers and engineers working in quantum photonics, quantum communications and cryogenic detection systems. We believe that iSNSPDs represent a rapidly advancing technology at the intersection of integrated photonics and quantum information processing systems. Substantial challenges remain in terms of scalability, reproducibility and performance trade-offs, continued progress along these directions is expected to push the limits of fully integrated photonic devices with active functionality for quantum applications.

\begin{backmatter}

\bmsection{Disclosures}
The authors declare no conflicts of interest.

\bmsection{Data availability} Data underlying the results presented in this paper are not publicly available at this time but may be obtained from the authors upon reasonable request.

\end{backmatter}

\bibliography{sample}

\end{document}